\documentclass[twocolumn, 5p]{elsarticle}

\usepackage{amsmath,amssymb}
\usepackage{balance}
\usepackage{subcaption}
\usepackage{graphicx}
\usepackage{hyperref}
\usepackage{multirow}
\usepackage{url}
\usepackage{color}
\usepackage{colortbl}
\usepackage{tabularx}
\usepackage{xspace}
\usepackage{booktabs}
\usepackage{listings}

\usepackage{algorithm}
\usepackage{algorithmicx}
\usepackage{amsmath,amssymb}
\usepackage{algpseudocode}

\usepackage{layouts}
\usepackage{wrapfig}

\usepackage{paralist}        

\usepackage{balance}          
\usepackage{multirow}         
\usepackage{multicol}     
\usepackage[font=small]{caption}         

\usepackage{comment}
\usepackage{siunitx}
\usepackage{hyperref}

\newcommand{\vct}[1]{\ensuremath{\boldsymbol{#1}}} 
\newcommand{\mat}[1]{\ensuremath{\mathbf{#1}}}
\newcommand{\set}[1]{\ensuremath{\mathcal{#1}}}

\newcommand{\T}{\ensuremath{\top}}

\newcommand{\argmax}{\operatornamewithlimits{\arg\,\max}}

\newcommand{\argmin}{\operatornamewithlimits{\arg\,\min}}

\newcommand{\ie}{{i.e.}\xspace}
\newcommand{\eg}{{e.g.}\xspace}
\newcommand{\etal}{{et al.}\xspace}
\newcommand{\etc}{{etc.}\xspace}

\newcommand{\ABC}{ABC\xspace}

\newcommand{\AdobeFlash}{Adobe Flash Player\xspace}

\newcommand{\ASVM}{\texttt{ASVM}\xspace}

\definecolor{dark-gray}{gray}{0.15}
\definecolor{black-gray}{gray}{0.2}
\definecolor{lightgray}{rgb}{.9,.9,.9}
\definecolor{darkgray}{rgb}{.4,.4,.4}
\definecolor{purple}{rgb}{0.8, 0.12, 0.52}
\definecolor{bluee}{rgb}{0.4, 0.4, 0.8}

\newcommand{\myparagraph}[1]{\smallskip \noindent \textbf{#1.}}

\lstdefinelanguage{ActionScript}{
	keywords={flash,utils,ByteArray,Endian,uint,readUnsignedInt,readUnsignedByte, LITTLE_ENDIAN, addCallback, RESPOND_PINGDATA, pingCallback, pingTimer, pingTimeouts, position, IG,
		submitForm,commitKey,event,doc,getAnnots,app,String,getField,isBoxChecked,checkThisBox,isBoxChecked,unescape,substring, Array,var,while,function,code,for,unescape,false,true,else,if, removeToolButton,cName,concat,addToolButton,cName,cEnable,cExec,nPage,return,util,printd, media, newPlayer 
	},
	keywordstyle=\color{purple}\bfseries,
	ndkeywords={class, export, boolean, throw, implements, import, this,null},
	ndkeywordstyle=\color{darkgray}\bfseries,
	identifierstyle=\color{black},
	sensitive=false,
	comment=[l]{//},
	morecomment=[s]{/*}{*/},
	commentstyle=\color{purple}\ttfamily,
	stringstyle=\color{bluee}\ttfamily,
	morestring=[b]',
	morestring=[b]"
}

\begin{document}

\begin{frontmatter}
	\title{Adversarial Detection of Flash Malware: Limitations and Open Issues}
	
	\address[1]{Department of Electrical and Electronic Engineering, University of Cagliari,
		Piazza d’Armi 09123, Cagliari, Italy}
	\address[2]{Pluribus One, Italy}
	\cortext[cor1]{Corresponding author}
	\fntext[fn1]{Those authors contributed equally to this manuscript.}
	\author[1]{Davide Maiorca\corref{cor1}\fnref{fn1}}
	\ead{davide.maiorca@unica.it}
	\author[1]{Ambra Demontis\fnref{fn1}}
	\ead{ambra.demontis@unica.it}
	\author[1,2]{Battista~Biggio}
	\ead{battista.biggio@unica.it}
	\author[1,2]{Fabio Roli}
	\ead{roli@unica.it}
	\author[1]{Giorgio Giacinto}

	\begin{abstract}
		During the past four years, Flash malware has become one of the most insidious threats to detect, with almost $600$ critical vulnerabilities targeting \AdobeFlash disclosed in the wild. Research has shown that machine learning can be successfully used to detect Flash malware by leveraging static analysis to extract information from the structure of the file or its bytecode.
        However, the robustness of Flash malware detectors against well-crafted evasion attempts - also known as adversarial examples - has never been investigated.
        In this paper, we propose a security evaluation of a novel, representative Flash detector that embeds a combination of the prominent, static features employed by state-of-the-art tools. In particular, we discuss how to craft adversarial Flash malware examples, showing that it suffices to manipulate the corresponding source malware samples slightly to evade detection.
        We then empirically demonstrate that popular defense techniques proposed to mitigate evasion attempts, including re-training on adversarial examples, may not always be sufficient to ensure robustness. 
        We argue that this occurs when the feature vectors extracted from adversarial examples become indistinguishable from those of benign data, meaning that the given feature representation is intrinsically vulnerable. In this respect, we are the first to formally define and quantitatively characterize this vulnerability, highlighting when an attack can be countered by solely improving the security of the learning algorithm, or when it requires also considering additional features.
        We conclude the paper by suggesting alternative research directions to improve the security of learning-based Flash malware detectors.
	\end{abstract}
	
	\begin{keyword}
		Adobe Flash \sep Malware Detection \sep Secure Machine Learning \sep Adversarial Training \sep Computer Security
	\end{keyword}
\end{frontmatter}

\section{Introduction}

\label{sect:introduction}

Malware detection is still a critical priority for researchers and industry, as countless of polymorphic attacks are continually being released~\cite{symantec19}.
In particular, a very insidious threat is represented by \emph{infection vectors}, \ie, files that exploit vulnerabilities of third-party applications to drop or execute malicious executables. Such vectors are typically documents (\eg, PDF, Office) or multimedia files (\eg Flash), and attackers leverage their structure to conceal scripting codes that exploit the target vulnerabilities.  For example, pages of PDF documents can hide malicious JavaScript codes or even additional executables~\cite{maiorca19-csur,maiorca19-sp}. 

To counteract such attacks, researchers proposed solutions based on machine-learning algorithms applied to information extracted by employing static or dynamic analysis of the embedded code, with excellent results for document-based infection vectors (\eg~\cite{maiorca12-mldm,smutz12-acsac,srndic13-ndss,smutz16-ndss}). However, research also showed that such detection approaches have several \emph{robustness} issues against evasion attacks, namely, well-crafted manipulations of the input samples at test time~\cite{biggio13-ecml,biggio14-tkde, biggio18} (also recently referred to as \emph{adversarial examples} in the context of deep learning~\cite{szegedy14-iclr,goodfellow15-iclr,grosse17-esorics}). The efficacy of such attacks depends on the information that the attacker possesses about the system, and on her ability to perform enough changes to the extracted features.

The robustness of machine learning approaches against adversarial attacks has been assessed in various case studies. In particular, some works employed code obfuscation through off-the-shelf tools or custom techniques~\cite{lin11-virology,singh16-virology,forse17-icissp,maiorca15-cose,scalas19-cose} to perform evasion. Other approaches directly targeted the learning algorithms by employing \emph{black-box} and \emph{white-box} evasion techniques (\eg, by using algorithms such as gradient-descent). Popular case-studies in which these techniques have been systematically studied are the detection of malicious PDF files~\cite{maiorca19-sp,xu16-ndss,srndic14,biggio13-ecml} and the detection of Android malware~\cite{demontis17-tdsc,yang17-acsac,calleja18-expert}. 

Multimedia malware, in particular in the Flash (also known as Adobe Small Web Format - SWF) format, has been studied considerably less from the perspective of adversarial attacks (in particular, against white- and black-box attacks). One of the reasons for such a lack of study is that, albeit Flash malware caused a massive uproar starting from $2015$ (due to the significant increment of vulnerabilities for Adobe Flash Player), its technology is going to be discontinued in $2020$. 
Nevertheless, this technology is still active from the perspective of malware-based attacks. Flash-based attacks represent excellent examples of highly obfuscated and evasive infection vectors. Critical vulnerabilities are still released for the Flash platform, and very dangerous attacks are still possible, such as the one that targeted North Korea in $2018$~\cite{trendmicro18}. 

We believe not only that it will take years before Flash completely stops being used, but also that analyzing Flash attacks constitutes an excellent test-bench for machine learning-based detection. The main reason for this claim is that most infection vectors (including the most used ones at the moment - \eg, PDF, Office documents, \etc) share a similar organization, composed of an external \emph{structure} and a code-based \emph{content}. Research on such infection vectors showed that the analysis of their structure and/or content led to encouraging results in their detection, and the Flash format is no exception in this~\cite{maiorca19-csur,wressnegger16-dimva,srndic16-eurasip}. Hence, the lessons learned from the study of adversarial attacks on formats like Flash can provide precious insights into the robustness of systems to detect other file formats.   

In this paper, we provide an in-depth analysis of Flash-based learning systems, focusing on their robustness against adversarial attacks. Our analysis aims to give interesting insights into how to design more secure systems for detecting infection vectors, and to point out possible vulnerabilities in the feature representations and the classification algorithms. 

To this end, we first propose a representative system for Flash malware detection, named FlashBuster. It is a static machine-learning system that employs information extracted by both the structure and the content of SWF files. Such an approach allows for a more comprehensive assessment of the extracted static information by representing and combining the contents employed by previous state-of-the-art systems~\cite{wressnegger16-dimva, srndic16-eurasip}.
We show that FlashBuster could detect the majority of malware samples in the wild, by obtaining comparable performances to other systems at state of the art, and demonstrate that it can predict previously unseen attacks. We also tested FlashBuster against popular obfuscation techniques, showing that our approach could also be employed in the presence of obfuscated malware.

We then evaluate FlashBuster robustness by simulating evasion attacks that leverage the knowledge that the attacker may possess about the targeted learning system~\cite{biggio14-tkde,biggio13-ecml, tramer16-usenix,xu16-ndss,papernot17-asiaccs,chen17-aisec,dang17-ccs}, against an increasing number of modifications to the input samples. 
The corresponding \emph{security evaluation curves}, depicting how the detection rate decreases against attack samples that are increasingly manipulated, allow us to understand and assess the vulnerabilities of FlashBuster under attack.

We finally discuss the effectiveness of \emph{adversarial training} against such evasive attacks.
To this end, we re-trained FlashBuster on the evasion attack samples used against it, and surprisingly show that this strategy may be ineffective in some cases. We argue that this is due to an \emph{intrinsic vulnerability} of the feature representation, \ie, to the fact that evasion attacks entirely mimic the feature values of benign data, thus becoming \emph{indistinguishable} for the learning algorithm. We define this vulnerability in formal terms, and quantitatively evaluate it by defining a specific metric that measures the extent to which the attack samples converge towards benign data. 

Our findings highlight a crucial problem that must be considered when designing secure machine-learning systems, \ie, that of evaluating \emph{in advance} the \emph{vulnerability of the given features}. Indeed, vulnerable information may compromise the whole system even if the employed decision function is robust.
In this respect, we sketch possible research directions that may lead one to design more secure machine learning-based malware detectors. 

Finally, we note that we publicly released FlashBuster, together with the features employed in the experiments for this paper.\footnote{\url{https://github.com/davidemaiorca/flashbuster}}

The rest of this paper is structured as follows. Section~\ref{sect:basics} provides the basics to understand the SWF format and an example of ActionScript code. Section~\ref{sect:relwork} describes the related work in the field. Section~\ref{sect:architecture} describes the architecture of FlashBuster. Section~\ref{sect:threat-model} describes the threat model in relation to adversarial environments. Section~\ref{sect:evasion-attacks} describes the evasion attacks employed in this paper. Section~\ref{sect:vuln} discusses the vulnerabilities that affect learning-based systems, and introduces a quantitative measure of feature and learning vulnerabilities. Section~\ref{sect:expeval} provides the experimental evaluation. Section~\ref{sect:overall} provides a discussion on the attained results. Section~\ref{sect:conclusions} closes the paper by sketching possible future work.

\section{ShockWave Flash File Format}
\label{sect:basics}
Small Web Format (SWF) is a file type that efficiently delivers multimedia contents, and it is processed by Adobe Software such as \AdobeFlash.\footnote{\url{https://get.adobe.com/en/flashplayer/}} 

SWF files are composed of three essential elements: \emph{(i)} a \emph{header} that describes important file properties such as the presence of compression, the version of the SWF format, and the number of video frames; \emph{(ii)} a list of \emph{tags}, i.e., data structures that establish and control the operations performed by the reader on the file data; \emph{(iii)} a unique tag called \texttt{End} that terminates the file. 

Some tags define actions such as pressing a button, moving the mouse, etc. These actions can be expanded further by employing a scripting language called ActionScript. ActionScript code (the latest release is $3.0$) is compiled to a bytecode that is run by the ActionScript Virtual Machine 2 (\ASVM $2$). The computation in the \ASVM $2$ is based on the execution of \emph{method bodies} composed of \emph{instructions}. Each method body runs in a specific \emph{context} that defines information such as default parameters. More about SWF and \ASVM $2$ can be found on the official SWF and VM references~\cite{adobe12-swf,adobe07-asvm}.

\subsection{ActionScript in SWF}
\label{sect:abcbasics:subsect:actionscript}

In order to better understand the role of ActionScript in SWF files, Listing \ref{sect:abcbasics:subsect:actionscript:listing:malw_example} shows a small snippet of code that is typically found in an ActionScript-based malware, where \texttt{ByteArray} structures are accessed. Such structures are employed by malware to store information about encrypted/decrypted URLs and payloads. 

The code in the Listing reads an \texttt{UnsignedByte} from the object \texttt{\_loc1\_}, which belongs to the class \texttt{IG} (which inherits from the \texttt{ByteArray} class - see the \texttt{coerce} bytecode instruction).  The code then performs a subtraction and assigns the output to the variable \texttt{isAS3}. This result will then be copied to another array of bytes (we did not report this action for space reasons).  Note how the reading is performed by following the \emph{little endian} (using the system-related \texttt{flash.utils.Endian} package) byte order. We point out that system API methods and classes are often essential for the attacker to build shellcodes or perform buffer overflows and heap spraying attacks. In fact, official ActionScript APIs allow managing low-level data structures efficiently, so attackers do not need to implement their memory management routines. 

From the bytecode perspective, to resolve correctly the package belonging to a specific method or class, the ActionScript Virtual Machine resorts to data structures called \texttt{Names}. Such structures are composed of one \emph{unqualified name} (for example, a class name) and one or more \emph{namespaces} that typically represent the \emph{packages} from which classes or methods are resolved. Normally, \texttt{Name} resolution occurs at compile time by associating one package to one class or method (\texttt{QName}). However, there are cases in which \texttt{Names} are resolved at runtime, in particular when there are multiple namespaces (packages) from which the same unqualified name (class) can be obtained. 

\begin{lstlisting}[language={ActionScript}, xleftmargin=0em, label=sect:abcbasics:subsect:actionscript:listing:malw_example, caption={An example of ActionScript. This code snippet is represented by its decompiled output (lines 2-5) and by its equivalent bytecode output (lines 11-25).},float=t]
_loc1_ = new IG();
_loc1_.endian = Endian.LITTLE_ENDIAN;
_loc1_.position = 0;
this.isAS3 = _loc1_.readUnsignedByte() - 1;


findpropstrict Qname(PackageNamespace(""),"IG")
constructprop Qname(PackageNamespace(""),"IG") 0
coerce Qname(PackageNamespace("flash.utils"),"ByteArray")
setlocal_1
getlocal_1
getlex Qname(PackageNamespace("flash.utils"),"Endian")
getproperty Qname(PackageNamespace(""),"LITTLE_ENDIAN")
setproperty Qname(PackageNamespace(""),"endian")
getlocal_1
pushbyte 0
setproperty Qname(PackageNamespace(""),"position")
getlocal_0
getlocal_1
callproperty Qname(PackageNamespace(""),"readUnsignedByte") 0
decrement
\end{lstlisting}

\section{Related Work}
\label{sect:relwork}

In the following, we provide a comprehensive review of the state-of-the-art approaches employed for Flash malware detection. Then, we also describe the prominent works in the field of adversarial machine learning.

\subsection{Flash Malware Detection}
Even if Flash-based malicious attacks started to grow considerably in $2015$, the number of detection approaches is rather limited.  
\texttt{FlashDetect}~\cite{overveldt12-raid} is one of the first approaches to the detection of ActionScript $3$-based malware. \texttt{FlashDetect} uses \texttt{Lightspark}, an open-source Flash viewer, to perform dynamic analysis of malicious Flash files. 
\texttt{FlashDetect} was employed inside the \texttt{Wepawet} service, which is not available anymore. 

\texttt{Gordon}~\cite{wressnegger16-dimva} is an approach that resorts to guided-code execution to detect malicious SWF files, by statically analyzing their ActionScript bytecode (without considering the file structure). In particular, the system selects the most suspicious security paths from the control flow graph of the code. Such paths usually have references to security-critical calls, such as the ones for dynamic code loading. To the best of our knowledge, \texttt{Gordon} is not publicly available. 

\texttt{Hidost}~\cite{srndic16-eurasip,hidost} is a static system that only focuses on the structure of the SWF file, without analyzing its ActionScript bytecode. More specifically, it considers sequences of objects belonging to the structure of the SWF file as features. The system evaluates the most occurring paths in the training dataset and extracts features based on the training data. Relying on such data might be dangerous from the perspective of targeted attacks, as a malicious test file with entirely different paths might evade detection.

According to the results in~\cite{wressnegger16-dimva,srndic16-eurasip}, both \texttt{Gordon} and \texttt{Hidost} performed well at detecting Flash malware. For this reason, we designed FlashBuster as a simplified extension of the aforementioned static systems, where information from both the structure and content of the file is extracted, and where the extracted features do not depend on the training distributions. 

Besides research-based prototypes, some off-the-shelf tools are often used to analyze SWF files, such as \texttt{JPEXS}~\cite{jpexs}, \texttt{PySWF}~\cite{pyswf}, \texttt{SWFReTools}~\cite{swfretools} and others. It is also possible to perform obfuscation of SWF files by employing \texttt{DoSWF}~\cite{doSWF}. In particular, this tool can either conceal variable names or introducing modifications to the flow graph that do not alter the overall semantics of the code. Additionally, it can perform full encryption of SWF files by dynamically loading them in memory at runtime. 

\begin{figure*}[t]
  \centering
  \includegraphics[width=0.65\textwidth]{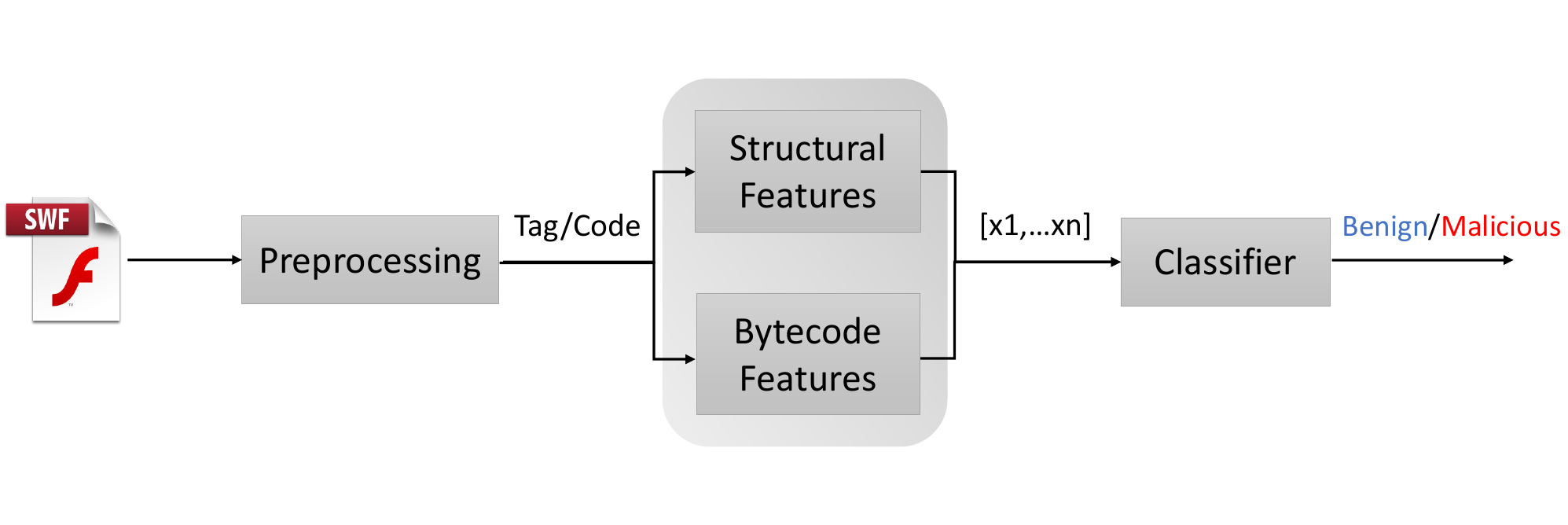}
  \caption{Graphical architecture of FlashBuster.}
  \label{sect:architecture:fig:system-outline}
\end{figure*}

\subsection{Adversarial Machine Learning} 

Several works doubted the security of machine learning since $2004$~\cite{biggio18}. The first two works in the field were proposed by Dalvi et al.~\cite{dalvi04} in 2004 and by Lowd and Meek~\cite{lowd05} in 2005. Those works, carried out in the field of spam filtering, demonstrated that it could be easy for the attackers to deceive classifiers at test time (\emph{evasion} attacks) by performing a limited amount of fine-grained changes to emails. 
Following works~\cite{barreno06-asiaccs,barreno10, biggio14-tkde} proposed attacker models and frameworks that are currently employed to assess the security of learning-based systems, also against training-time (\emph{poisoning}) attacks. 
The first gradient-based evasion~\cite{biggio13-ecml} and poisoning~\cite{biggio12-icml} attacks were proposed by Biggio et al. respectively in the 2013 and 2012. In~\cite{biggio13-ecml}, the authors introduced two important concepts that are currently adopted in the adversarial field, namely \emph{high-confidence} adversarial examples and the use of a \emph{surrogate} model. The over-mentioned works anticipated the discovery of the so-called \emph{adversarial examples} against deep neural networks~\cite{szegedy14-iclr,goodfellow15-iclr}.
\newline
The vulnerability to evasion attacks has been then especially studied on learning systems designed to detect malware samples (for example, on PDF files~\cite{maiorca19-csur,xu17-usenix, xu16-ndss,srndic14-sp, maiorca13-asiaccs}), thus raising serious concerns about their usability under adversarial environments. These assessments have been recently expanded to systems for Android malware detection~\cite{demontis17-tdsc,yang17-acsac,calleja18-expert,melis18-eusipco,scalas19-cose} and to binary executables (with the adoption of deep learning)~\cite{raff17-arxiv,kolosnjaji18-eusipco}.

\section{FlashBuster Architecture}
\label{sect:architecture}
FlashBuster is a static, machine learning-based system whose aim is to detect malicious SWF files and to distinguish them from benign ones.
Its main goal is reproducing a combination of the characteristics of previous state-of-the-art systems (see Section \ref{sect:relwork}) that proved to be effective in detecting SWF malware, in order to assess the efficacy and robustness of the information extracted to recognize these attacks. To this end, FlashBuster leverages information provided by the tag structure and the ActionScript bytecode of the file. Figure \ref{sect:architecture:fig:system-outline} shows the general architecture of the system, which can be divided into three modules:

\myparagraph{Parser} This module analyzes the SWF file and extracts information about its structure and its ActionScript bytecode.

\myparagraph{Feature Extractor} This module transforms the information obtained from the parser in a vector of numbers, which characterizes the whole SWF file. 

\myparagraph{Classifier} This module decides on the maliciousness of the SWF file according to the feature vector it receives as input. Such a module is a mathematical function that tunes its parameters by receiving various examples taken from a so-called \emph{training set}. Once its parameters have been set up, the classifier can recognize malicious files that have not been included in the training examples.

In the following, we provide a more detailed description of each component of the system.

\subsection{Parser}
\label{sect:architecture:subsect:parser}
As previously said, this module performs data pre-processing and selects the information that will be further processed by the other modules.
FlashBuster leverages a modified version of \texttt{JPEXS}, a powerful, Java-based Flash disassembler and decompiler~\cite{jpexs}. This software is based on \texttt{RABCDasm}~\cite{rabcdasm}, one of the most popular Flash disassemblers, and it adds new features such as de-obfuscation, file debugging and preview, etc.

More specifically, the parser featured by FlashBuster executes the following operations:
\emph{(i)} it performs static de-obfuscation of ActionScript code. This operation is important, as some malicious files might use name obfuscation or other popular techniques to conceal attacks;
\emph{(ii)}  it extracts the complete SWF structure in terms of \emph{tags};
\emph{(iii)} it disassembles the \ABC bytecode so that it could be read as a plain-text file. This operation includes automatic de-obfuscation of the ActionScript code. Both the tag structure and the \ABC bytecode are sent to the feature extractor module for further analysis. 

\subsection{Feature Extraction}
\label{sect:architecture:subsect:features}
This module represents the core of the whole system, and it converts the information extracted by the parser to a vector of numbers that will be sent to the classifier. Our goal here was not devising a completely novel feature set, but proposing a comprehensive approach that could be comparable, in terms of detection performances, to other state-of-the-art approaches. For this reason, we referred to~\cite{wressnegger16-dimva} to implement a conceptually similar system that would employ information extracted from the structure and the content of the SWF file. However, differently to~\cite{wressnegger16-dimva}, we extracted the \emph{number of occurrences} of the extracted information, instead of analyzing their sequences. This intuition derives from the experience in PDF malware detection~\cite{maiorca19-csur}, where the occurrences of information extracted from the structure and the content of a PDF file proved to be greatly effective in performing detection. We now provide a detailed description of the features extracted in the two cases. 

\subsubsection{Structural Features (Tags)}
\label{sect:architecture:subsect:features:subsubsect:structure}
These features are related to the information that can be extracted from SWF tags and are crucial to understanding which objects and actions are executed by the SWF file. 
The main idea here is that malware does not contain particularly complex multimedia contents, such as video with many frames or audio files. Various malware samples simply display images such as rectangles or blank backgrounds. For this reason, we extracted the following $14$ features from the file structure, corresponding to the \emph{number of occurrences} of specific SWF tags within the file\footnote{The reader can find more information about these tags on the official SWF specification \cite{adobe12-swf}.}: 

\myparagraph{Frames} \texttt{ShowFrame} tags that are used to display frames.

\myparagraph{Shapes} \texttt{DefineShape} (in any of all its four variants) tags, used to define new shapes that will be plotted on the screen.

\myparagraph{Sounds} Sound-related events, extracted by examining any of the following tags: \texttt{DefineSound}, \texttt{SoundStreamHead1}, 
\texttt{SoundStreamHead-2} and \texttt{SoundStreamBlock}.

\myparagraph{BinaryData} Groups of embedded data, represented by the tag \texttt{DefineBinaryData}.

\myparagraph{Scripts} ActionScript codes that are contained in the file. Note that a SWF file does not necessarily require scripting code to perform its operations, especially in benign files (ActionScript has been initially devised as an aid to the execution of SWF files). Scripts are discovered by analyzing the following tags: \texttt{DoABC}, \texttt{DoABCDefine}, \texttt{DoInitAction}, \texttt{DoAction}.

\myparagraph{Fonts} Font-related objects, extracted by detecting any of the following tags: \texttt{DefineFont} (in all its variants), \texttt{DefineCompactedFont}, \texttt{DefineFontInfo} (in all its variants), \texttt{DefineFontName}. 

\myparagraph{Sprites} Sprites extracted by examining the tag \texttt{DefineSprite}.

\myparagraph{MorphShapes} Morphed shapes (i.e., shapes that might transform into new ones) extracted by examining the tag \texttt{DefineMorphShape} (and its variants).

\myparagraph{Texts} Text-related objects, extracted by checking any of the following tags: \texttt{DefineText} (along with  its variants) and \texttt{DefineEditText}.

\myparagraph{Images} Images contained in the file, extracted by examining any of the following tags (and their variants): \texttt{DefineBits}, \texttt{DefineBitsJPEG}, \texttt{JPEGTables} and \texttt{DefineBitsLossless}.

\myparagraph{Videos} The number of embedded videos in the file, extracted by examining the tags \texttt{DefineVideoStream} and \texttt{VideoFrame}.  

\myparagraph{Buttons} Buttons with which the user can interact with the SWF content. Such counting can be done by examining the tag \texttt{DefineButton} (along with its variants).

\myparagraph{Errors} Errors made by the parser when analyzing specific tags. Such errors often occur, for example, when the SWF file is malformed due to errors in its compression. 

\myparagraph{Unknown} Tags that do not belong to the SWF specifications (probably malformed tags).

It is worth noting that we preferred counting the occurrences of each tag (instead of just considering its presence/absence) because we observed that benign objects contain significantly more tags of the same type in comparison to malicious files. 
We also observe that structural features must be carefully treated, as benign and malicious files can be similar to each other concerning their tag structure. For this reason, structural features alone are not enough to ensure reliable detection and must be integrated with information from the scripted content of the file. 

\subsubsection{Actionscript Bytecode Features (API calls)}
\label{sect:architecture:subsect:features:subsubsect:bytecode}
As structural features (\ie, tags) might suffer from the limitations mentioned in Sect.~\ref{sect:architecture:subsect:features:subsubsect:structure}, we employed an additional set of features that focus on the content of the scripting code that might be included in the files.  Although it is not strictly necessary to use ActionScript for benign operations, its role is essential to execute attacks. In particular, as shown in Sect.~\ref{sect:abcbasics:subsect:actionscript}, the attacker usually needs to resort to system APIs to perform memory manipulation or to trigger specific events. Moreover, APIs can be used to communicate with external interfaces or to contact an external \texttt{URL} to automatically drop malicious content on the victim's system.  

System APIs belong to the official Adobe ActionScript specifications~\cite{as_ref}. For this reason, we created an additional feature set that counts the classes and methods belonging to such specifications, leading to $4724$ new features. More specifically, this feature set represents the number of specific System methods and classes inside the bytecode. We chose to use only system-based APIs for two reasons: ($i$) the feature vector does not include user-defined APIs, so that the feature list is independent of the training data that is considered for the analysis; ($ii$) system-based calls are more difficult to obfuscate, as the user does not directly implement them. 

 With respect to the example described in Sect.~\ref*{sect:abcbasics:subsect:actionscript}, we would therefore consider as valid features the classes \texttt{flash.utils.ByteArray} and \texttt{flash.utils.Endian}, and the method \texttt{readUnsignedByte}. On the contrary, we would not directly consider the class name \texttt{IG}, as it was directly implemented by the user. 
  The rationale behind \emph{counting} the occurrences of system-based methods and classes is that an attacker might systematically use functions to manipulate memory or perform suspicious actions. Alternatively, she might attempt to trigger events repeatedly or to access specific interfaces. 
 
 Finally, we observe that all features were normalized with the popular tf-idf strategy~\cite{baeza99}. This normalization is particularly crucial for SVM classifiers, which typically perform best with normalized features.

\subsection{Classification}
\label{sect:architecture:subsect:classification}

The features extracted with FlashBuster can be used with different classification algorithms. In the experimental evaluation that we describe in Section \ref{sect:expeval}, we report the results for different classification algorithms. In particular, we focused our attention on SVM (linear and non-linear) and Random Forests, as these were successfully employed in other works on SWF detection (\eg,~\cite{wressnegger16-dimva}). Although other classifiers (or even their ensembles) may be employed, we believe that the chosen set is representative of the majority of classifiers in the wild and that similar results would hold for possible alternatives to classification~\cite{maiorca19-sp}.

\section{Attack Model}
\label{sect:threat-model}
To assess the security of FlashBuster against adversarial manipulation of the input data (which can be either performed at training time or at test time), we leverage an attack model originally defined in the area of adversarial machine learning~\cite{biggio18,biggio14-tkde,biggio14-ijprai}. It builds on the well-known taxonomy of Barreno~\etal~\cite{barreno06-asiaccs,barreno10,huang11} which categorizes potential attacks against machine-learning algorithms along three axes: \emph{security violation},  \emph{attack specificity} and  \emph{attack influence}. By exploiting this taxonomy, the attack model enables defining various potential attack scenarios, in terms of explicit assumptions on the attacker's goal, knowledge of the system, and capability of manipulating the input data.

\subsection{Notation}
\label{sect:threat-model:subsect:notation}
For the sake of clarity, we report here a table of the symbols that will be used during the rest of the paper, along with a brief description. Such a table can be employed as a reference during the reading of the following sections.

\begin{table}[!h]
	\centering
	\caption{Notation and symbols employed in the paper.}
	\label{sect:threat-model:subsect:notation:tab:families}
	\resizebox{0.45\textwidth}{!}{%
		\begin{tabular}
			{ |>{\bfseries}l  |l| }
			\hline
			\rowcolor[rgb]{.9,.9,1} \textbf{Symbol} &  \textbf{Description} \\
			\hline
            \textbf{$\set W$ } & Objective function \\ \hline
            \textbf{$\set A$ and $\set A^{\prime}$} & Input samples and modified attack samples \\ \hline
            \textbf{$\vct \theta$} & System knowledge \\ \hline
            \textbf{$\set D$ and $\hat{\set D}$ } & Training data and surrogate training\\ \hline
            \textbf{$\set X$} & Feature set\\ \hline
            \textbf{$\set L$} & Learning algorithm \\ \hline
            \textbf{$f$, $\hat{f}$} & Decision function, surrogate classifier\\ \hline
            \textbf{$f\star$} & Classifier with minimum Bayesian error\\ \hline
            \textbf{$\Theta$} & Space that embeds $\set D$, $\set X$, $\set L$, $f$  \\ \hline
            \textbf{$\Omega$} & Set of possible modifications \\ \hline
            \textbf{$\Phi$} & Feature extraction function \\ \hline
            \textbf{$\vct x$ and $\vct x^{\prime}$} & Original and modified feature vectors \\ \hline
            \textbf{$\vct z$, $\vct z^{\prime}$ and $\vct z^{\star}$} & Original, modified, and optimal input samples\\ \hline
            \textbf{$\vct x_{\rm lb}$ and $\vct x_{\rm ub}$} & box-constraint bounds\\ \hline
            \textbf{$\varepsilon$} & Number of feature modifications\\ \hline
            \textbf{$y$} & Labels\\ \hline
            \textbf{$E$} & Classification error\\ \hline
            \textbf{$\ell$} & Zero-one loss\\ \hline
            \textbf{$p$} & Data distribution\\ \hline
            \textbf{$a$} & Deviation from data distribution\\ \hline
            \textbf{$E$} & Classification error\\ \hline
            \textbf{$BC$} & Bhattacharyya Coefficient\\ \hline
            \textbf{$\mu$, $\Sigma_b$} & Mean and covariance of data\\ \hline

		\end{tabular}}
\end{table}

\subsection{Attacker's Goal} 
\label{sect:threat-model:subsect:goal}
It is defined in terms of two characteristics, \ie, security violation and attack specificity.

\myparagraph{Security violation} In security engineering, a system can be violated by compromising its \emph{integrity}, \emph{availability}, or \emph{confidentiality}. Violating the integrity of FlashBuster amounts to having malware samples undetected; its \emph{availability} is compromised if it misclassifies benign samples as malware, causing a denial of service to legitimate users; and \emph{confidentiality} is violated if it leaks confidential information about its users.

\myparagraph{Attack specificity} The specificity of the attack can be \emph{targeted} or \emph{indiscriminate}, based on whether the attacker aims to have only specific samples misclassified (\eg, a specific malware sample to infect a particular device or user), or if any misclassified sample meets her goal (\eg, if the goal is to launch an indiscriminate attack campaign).

We formalize the attacker's goal here in terms of an objective function $\set W (\set A^{\prime}, \vct \theta) \in \mathbb R$ (where $\vct \theta$ is the knowledge possessed by the attacker about the system), which evaluates to what extent the manipulated attack samples $\set A^{\prime}$ meet the attacker's goal.

\subsection{Attacker's Knowledge}
\label{sect:threat-model:subsect:knowledge}
The attacker may have different levels of knowledge of the targeted system~\cite{biggio18,biggio14-tkde,biggio14-ijprai,barreno06-asiaccs,barreno10,huang11,srndic14}. In particular, she may know completely, partially, or do not have any information at all about:
($i$) the training data $\set D$;
($ii$) the feature set $\set X$, \ie, how input data is mapped onto a vector of feature values;
($iii$) the learning algorithm $\set L(\set D, f)$, and its decision function $f(\vct x)$, including its (trained) parameters (\eg, feature weights and bias in linear classifiers), if any. In some applications, the attacker may also exploit feedback on the classifier's decisions to improve her knowledge of the system, and, more generally, her attack strategy~\cite{biggio18,biggio14-tkde,biggio14-ijprai,barreno06-asiaccs,huang11}.

The attacker's knowledge can be represented in terms of a space $\Theta$ that encodes knowledge of the data $\set D$, the feature space $\set X$, the learning algorithm $\set L(\set D, f)$ and its decision function $f$. 
In particular, we distinguish between limited- and perfect-knowledge attacks.

\subsubsection{Limited-Knowledge (LK) Black-Box Attacks}
Under this scenario, the attacker is typically only assumed to know the feature representation $\set X$ and the learning algorithm $\set L$, but not the training data $\set D$ and the trained classifier $f$.
This assumption is common under the security-by-design paradigm: the goal is to show that the system may be reasonably secure even if the attacker knows how it works but does not know any detail on the specific deployed instance~\cite{biggio14-tkde,biggio14-ijprai,barreno06-asiaccs,huang11,biggio13-ecml,demontis17-tdsc,biggio18}.

In particular, according to the definition proposed by Biggio et al., we distinguish the cases in which either the training data or the trained classifier are unknown \cite{biggio17-aisec}. In the first case, to which we refer as LK attacks with \emph{surrogate data}, it is often assumed that the attacker can collect a surrogate dataset $\hat{\set D}$ and that she can learn a surrogate classifier $\hat f$ on $\hat{\set D}$ to approximate the true $f$~\cite{biggio13-ecml,papernot17-asiaccs}. Note also that the class labels of $\hat{\set D}$ can be modified using the feedback provided from the targeted classifier $f$, when available (e.g., as an online service providing class labels to the input data). The knowledge-parameter vector can be thus encoded as $\vct \theta_{LK-SD} = (\hat{\set D}, \set X, \set L, \hat f)$.

In the second case, to which we refer to as LK attacks with \emph{surrogate learners}, we assume that the attacker knows the training distribution ${\set D}$, but not the learning model. Hence, she trains a surrogate function on the same training data. Hence, the knowledge-parameter vector can be encoded as $\vct \theta_{LK-SL} = (\set D, \set X, \hat{\set L}, \hat f)$.

\subsubsection{Perfect-Knowledge (PK) White-Box Attacks}

This is the worst-case setting in which also the targeted classifier is fully known to the attacker, \ie, $\vct \theta = (\set D, \set X, \set L, f)$. Although it is not very likely to happen in practice that the attacker gets to know even the trained classifier's parameters, this white-box setting is particularly interesting as it provides an upper bound on the performance degradation incurred by the system under attack, and can be used as a reference to evaluate the effectiveness of the system against the other (less pessimistic) attack scenarios. In the experimental evaluation of this work, we will mostly explore this knowledge scenario, along with limited knowledge with surrogate learners.

\subsection{Attacker's Capability} \label{sect:threat-model:subsect:capability} 
The attacker's capability of manipulating the input data is defined in terms of the so-called \emph{attack influence} and it is based on some application-specific constraints.

\myparagraph{Attack Influence} This defines whether the attacker can only manipulate data at test time (\emph{exploratory} influence), or if she can also contaminate the training data (\emph{causative} influence). Such contamination is possible, for instance, if the system is retrained online using data collected during operations that can be manipulated by the attacker~\cite{barreno06-asiaccs,huang11,biggio14-tkde, biggio18}.

\myparagraph{Application-specific constraints} According to the given application, these constraints define how and to which extent the input data (and its features) can be modified to reach the attacker's goal. 
In many cases, these constraints can be directly encoded in terms of distances in the feature space, computed between the source malware data and its manipulated versions~\cite{dalvi04,lowd05,globerson06-icml,teo08,bruckner12,biggio13-ecml, biggio18}. 
FlashBuster is not an exception to this rule, as we will discuss in the remainder of this section.
In general, the attacker's capability can thus be represented in terms of a set of possible modifications $\Omega(\set A)$ performed on the input samples $\set A$.

\subsection{Attack Strategy} 
\label{sect:threat-model:subsect:strategy}
The attack strategy amounts to formalizing the derivation of the attack in terms of an optimization problem~\cite{biggio13-ecml,biggio14-tkde}.
Given the attacker's goal $\set W (\set A^{\prime}, \vct \theta)$, along with a knowledge-parameter vector $\vct \theta\in\Theta$ and a set of manipulated attacks $\set A^{\prime} \in \Omega(\set A)$, 
the attack strategy is given as:

\begin{equation}
\set A^{\star} =\textstyle \argmax_{\set A'\in\Omega(\set A)} \, \set W(\set A'; \vct \theta) \,.
\label{eq:opt-attack}
\end{equation}
Under this formulation, one can characterize different attack scenarios. The two main ones often considered in adversarial machine learning are referred to as classifier {\bf evasion} and {\bf poisoning}~\cite{biggio13-ecml,biggio12-icml,mei15-aaai,biggio17-aisec,barreno06-asiaccs,barreno10,huang11,biggio14-tkde,biggio14-ijprai, biggio18}.
In the remainder of this work we focus on \emph{classifier evasion}, while we refer the reader to~\cite{biggio14-tkde,mei15-aaai,biggio17-aisec} for further details on \emph{classifier poisoning}.

\section{Evasion Attacks and Security Scenarios} \label{sect:evasion-attacks}

Evasion attacks consist of manipulating malicious samples at test time to have them misclassified as benign by a trained classifier. The attacker's goal is thus to violate system \emph{integrity}, either with a \emph{targeted} or with an \emph{indiscriminate} attack, depending on whether the attacker is targeting a specific machine or running an indiscriminate attack campaign.
More formally, evasion attacks can be written in terms of the following optimization problem:
\begin{equation}
\vct z^{\star} = \argmin_{\vct z^{\prime} \in \Omega( \vct z )} \hat f(\Phi(\vct z^{\prime}))  \, ,
\label{eq:evasion}
\end{equation}
where $\vct x^{\prime} = \Phi(\vct z^{\prime})$ is the feature vector associated to the modified attack sample $\vct z^{\prime}$, $\vct x = \Phi(\vct z)$ is the feature vector associated to the source (unmodified) malware sample $\vct z$, $\Phi$ is the feature extraction function, and $\hat{f}$ is the surrogate classifier estimated by the attacker.
Concerning Eq.~\eqref{eq:opt-attack}, note that here samples can be optimized one at a time, as they can be independently modified.

As in previous work~\cite{biggio14-tkde,biggio13-ecml,demontis17-tdsc}, we first simulate the attack at the feature level, \ie, we directly manipulate the feature values of malicious samples without constructing the corresponding real-world samples while running the attack. We discuss in Sect.~\ref{sect:evasion-attacks:subsect:constr-adv-ex} how to create the corresponding real-world evasive malware samples.
The above problem can be thus simplified as:
\begin{eqnarray}
\label{eq:ev1}
\vct x^* = && \textstyle \argmin_{\vct x^{\prime}}  \hat f(\vct x^{\prime} ) \\
\label{eq:ev2}
{\rm s. t.} && \| \vct x^{\prime} - \vct x \|_1 \leq \varepsilon \; , \\ 
\label{eq:ev3}
&& \vct x_{\rm lb} \preceq \vct x^{\prime} \preceq \vct x_{\rm ub} \, ,
\end{eqnarray}
where we have also made the manipulation constraints $\Omega$ used to attack FlashBuster explicit.
In particular, the box constraint $\vct x_{\rm lb} \preceq \vct x^{\prime} \preceq \vct x_{\rm ub}$ (in which the inequality holds for each element of the vector) bounds the minimum and maximum feature values for the attack sample $\vct x^{\prime}$. For FlashBuster, we will only consider \emph{feature injection}, \ie, we will only allow the injection of structural and bytecode features within the SWF file to avoid compromising the intrusive functionality of the malware samples (something that can easily happen by deleting objects or specific calls). This can be simply accounted for by setting $\vct x_{\rm lb} = \vct x$.
The additional $\ell_1$ distance constraint $\| \vct x^{\prime} - \vct x \|_1 \leq \varepsilon$ thus sets the maximum number $\varepsilon$ of structural and bytecode features (\ie, tags and API calls) that can be injected into the file.
The solution to the above optimization problem amounts to identifying which features should be modified to maximally decrease the value of the classification function, \ie, to maximize the probability of evading detection~\cite{biggio14-tkde,biggio13-ecml}. This set of features varies depending on the input sample $\vct x$.

\begin{algorithm}[t]
	\caption{Evasion Attack}
	\label{alg:evasion}
	\begin{algorithmic}[1]
		\Require $\vct x$, the malicious sample;  $\vct x^{(0)}$, the initial location of the attack sample;
		$\hat f$, the surrogate classifier (Eq.~\ref{eq:ev1});
		$\varepsilon$, the maximum number of injected structural and bytecode features (Eq.~\ref{eq:ev2});
		$\vct x_{\rm lb}$ and $\vct x_{\rm ub}$, the box constraint bounds (Eq.~\ref{eq:ev2});
		$\epsilon$, a small positive constant.
		\Ensure $\vct x^{\prime}$, the evasion attack sample.
		\State $i \gets 0$
		\Repeat
		\State $i \gets i + 1$
		\State $t^{\prime} = \argmin_{t} \hat f(\Pi (\vct x^{(i-1)} - t \nabla \hat f(\vct x^{(i-1)}) ))$
		\State $\vct x^{(i)} \gets \Pi( \vct x^{(i-1)} - t^{\prime} \nabla \hat f(\vct x^{(i-1)}))$ 
		\Until{$ | \hat f(\vct x^{(i)}) - \hat f(\vct x^{(i-1)}) | < \epsilon$}
		\State \Return $\vct x^{(i)} $
	\end{algorithmic}
\end{algorithm}

\subsection{Evasion Attack Algorithm}
\label{sect:evasion-attacks:subsect:algorithms}
If the objective function (\ie, the decision function of the classifier) $f$ is not linear, as for kernelized SVMs and random forests, Problem \eqref{eq:ev1}-\eqref{eq:ev2} corresponds to a non-linear programming problem with linear constraints. The solution is, therefore, typically found at a local minimum of the objective function.
Problem \eqref{eq:ev1}-\eqref{eq:ev2} can be solved with standard algorithms, but this is not typically very efficient, as such solvers do not exploit specific knowledge about the evasion problem. We thus devise an ad-hoc solver based on exploring a descent direction aligned with the gradient $\nabla \hat f(\vct x^\prime)$ using a bisect line search, similar to that used in our previous work~\cite{russu16-aisec}. Its basic structure is given as Algorithm~\ref{alg:evasion}. To minimize the number of iterations, we explore one feature at a time (starting from the most promising feature, \ie, the one exhibiting the highest gradient variation in absolute value), leveraging the fact that the solution will be sparse (as the problem is $\ell_1$ constrained). We also minimize the number of gradient and function evaluations to further speed up our evasion algorithm; \eg, we only re-compute the gradient of $\hat f(\vct x)$ when no better point is found on the direction under exploration. Finally, we initialize $\vct x^{(0)}$ twice (first starting from $\vct x$, and then from a benign sample projected onto the feasible domain), to mitigate the problem of ending up in a local minimum that does not evade detection.\footnote{This problem has been first pointed out in~\cite{biggio13-ecml}, where the authors have introduced a \emph{mimicry} term to overcome it. Here we consider a different initialization mechanism, which allows us to get rid of the complicated mimicry term in the objective function.}
 
 \subsection{Constructing Adversarial Malware Examples}
 \label{sect:evasion-attacks:subsect:constr-adv-ex}
 
A common problem when performing adversarial attacks against machine learning is evaluating whether they can be truly performed \emph{in practice}. As gradient-descent attacks are performed at the feature level, the attacker is then supposed to solve the so-called \emph{inverse feature-mapping problem}, \ie, to reconstruct from the obtained features the sample that can be concretely used against the target classifier~\cite{huang11,biggio14-tkde,demontis17-tdsc}.

Such an operation is not smooth to perform in many cases, not only from a more theoretical standpoint (as discussed in~\cite{huang11}) but also from a practical perspective.
In the specific case of Flash malware (as well as malware in general), generating the corresponding real-world adversarial examples may be complicated, as a single wrong operation can compromise the intrusive functionality of the embedded exploitation code~\cite{demontis17-tdsc}.
For example, removing one structural feature such as one frame or script might entirely break the SWF file. For this reason, in this paper, we only considered \emph{injection} of additional content into the SWF file. In particular, we propose a methodology to automatically construct the real evasive samples by automatically injecting features that do not alter the overall functionality of the SWF files. This methodology applies to content-based features, which represent the majority of features employed by FlashBuster, and it works as follows:

\begin{enumerate}
    \item We disassemble the target SWF file by extracting its ActionScript bytecode (by using, \eg, tools such as \texttt{RABCDasm}).
    \item We explore the disassembled code until we find \texttt{return}-type instructions in functions (\eg, \texttt{returnvoid}).
    \item We inject a set of instructions that are never parsed by the code (as it comes after the return instruction). In particular, we inject a combination of two instructions: \texttt{pushstring}, combined with the name of the API call (\eg, \texttt{pushstring "flash.utils.ByteArray"}); \texttt{pop}, which essentially removes the string pushed into the stack. This combination does not alter the memory of the virtual machine.
    \item The disassembled code is reassembled back to the original file.
\end{enumerate}

With this technique, it is possible to increase the content-based feature values of FlashBuster without changing the behavior of the file. We tested this methodology on various samples by using the sandbox-based analyzer \texttt{Any.run}~\cite{anyrun}\footnote{It is not possible to use \texttt{Any.run} to perform analysis of groups of samples. Each sample should be analyzed singularly through the graphical interface.}. In particular, we run the samples before and after the reconstruction, by comparing the extracted reports. In the tested cases, we found no differences in behavior between the original and the modified files. 

It is worth noting that unreachable code injected after \texttt{return} instructions could be discarded by making FlashBuster analyze the applications flow-graph. We plan to implement control flow-based parsing in future work. However, it would still be possible for an attacker to inject unreachable code while defying control flow analysis. For example, attackers may employ \emph{opaque predicates}, which are conditional branches that are typically extremely hard to evaluate statically~\cite{collberg98-sigact}. In this case, unreachable code can be added as an apparently legitimate branch that is never taken, and advanced static (or dynamic) analysis would be required to detect and discard it.    

A similar strategy can be employed to modify structural-based features. By using libraries such as \texttt{PySWF}~\cite{pyswf}, it is possible to extract the structures of tags and add new ones. However, changing the number of tags alters the whole SWF structure, which must be then reconstructed with the correct offsets. This operation is not easy to carry out in an automatic fashion. However, it is possible to use \texttt{JPEXS} to edit and reconstruct the SWF structure manually. We plan, in future work, to extend the automatic reconstruction of samples also to structural.

\subsection{Summary of the Evasion Methodology}
\label{sect:evasion-attacks:subsect:summary}
As a further clarification of what we described in Sections~\ref{sect:threat-model} and ~\ref{sect:evasion-attacks}, we provide a brief, practical summary of the attack scenario and the evasion methodology that will be employed in the following of this paper.
\begin{itemize}
    \item The goal of the attacker is to modify Flash applications classified as malicious by the target system to make them misclassified as benign.
    \item The attacker possesses full knowledge of the target system. She knows which features were used to train the model, the classifier, and the trained model itself (Perfect Knowledge Scenario). The only exception is represented by attacks against non-differentiable Random Forest classifiers, where the attacker employs a differentiable surrogate model to perform the attack (Limited Knowledge with surrogate models).
    \item The attacker performs evasion against the trained model by employing a gradient descent algorithm. The role of gradient descent is selecting which features should be changed to maximize the probability of evasion with the minimum number of changes to the application. The attack is performed on the feature level and outputs a modified feature vector.
    \item Once the attacker manages to attain evasion for a specific sample, she can concretely inject the features to the sample in order to obtain the corresponding evasive feature vector. 
\end{itemize}

\section{Feature and Learning Vulnerability}
\label{sect:vuln}

\begin{figure}[t]
\begin{center}
\includegraphics[width=0.48\textwidth]{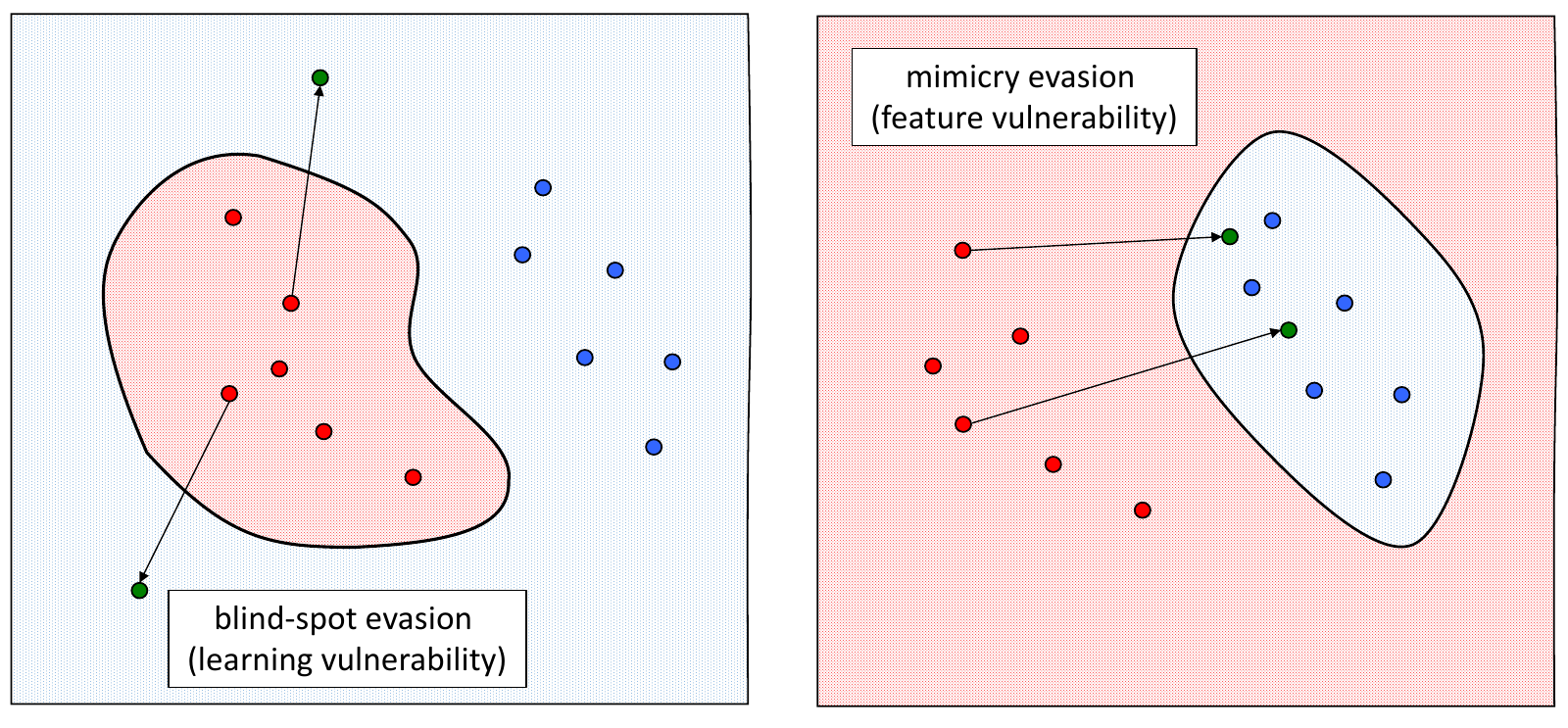}
\caption{Conceptual representation of \emph{learning} (left) and \emph{feature} (right) vulnerability. Red, blue and green samples represent, respectively, malicious, benign and attack samples.}
\label{fig:vuln}
\end{center}
\end{figure}

We discuss here an interesting aspect related to the vulnerability of learning-based systems, first highlighted in~\cite{biggio15-mcs,russu16-aisec}, and conceptually represented in Fig.~\ref{fig:vuln}, which shows two classifiers on a two-feature space. The classifiers can be defined as surfaces closed around malicious (left) and benign (right) samples. The red, blue, and green samples represent, respectively, malicious, benign, and attack samples. 
An evasion attack sample is typically misclassified in two scenarios: ($i$) the feature vector related to the sample is \emph{far enough} from those belonging to the rest of known training samples (both malicious and benign), or ($ii$) the feature vector is \emph{indistinguishable} from those exhibited by benign data.

In the first scenario, usually referred to as \emph{blind-spot evasion}, retraining the classifier on the adversarial examples (with \emph{adversarial training}) should successfully enable their detection, improving classifier security. 
This means that the classification error induced by such attacks could be \emph{reduced} in advance, by designing a learning algorithm capable of anticipating this threat; \eg, building a classifier that better encloses benign data and classifies as malicious the regions of the feature space where training data is absent or scarce (see, \eg, the classifier in the right plot of Fig.~\ref{fig:vuln}).
We refer to this vulnerability as one induced by the \emph{learning algorithm} (left plot in Fig.~\ref{fig:vuln}).

In the second scenario, instead, retraining the classifier would be useless, as the whole distribution of the evasion samples is overlapped with that of benign data in feature space, \ie, the attack increases the Bayesian (non-reducible) error. We thus refer to this attack as \emph{mimicry evasion}, and the corresponding vulnerability as one induced by the \emph{feature representation} (right plot in Fig.~\ref{fig:vuln}).
In fact, if a malware sample can be modified to exhibit the same feature values of benign data, it means that the given features are intrinsically weak, and no secure learning algorithm can prevent this issue.

This notion can also be motivated in formal terms, similarly to the risk analysis reported in~\cite{biggio15-mcs}. 
From a Bayesian perspective, learning algorithms assume an underlying (though unknown) distribution $p(\vct x, y)$ governing the generation of benign ($y=-1$) and malicious ($y=+1$) data, and aim to minimize the classification error $E(f) = \mathbb{E}_{(\vct x, y) \sim p } \ell( y, f(\vct x))$, where $\mathbb{E}$ is the expectation operator, $\ell$ is the zero-one loss, and $f$ is the classification function returning the predicted class label (\ie, $\pm 1$).
Let us denote the optimal classifier achieving the minimum (Bayesian) error on $p$ with $f^\star$.
It is clear that, if there is no \emph{evidence} $p(\vct x)$ of (training) data in some regions of the feature space (usually referred to as \emph{blind spots}), such regions can be arbitrarily classified by $f^\star$ as either benign or malicious with no impact on the classification error (the expectation on $p$ will be in any case zero in those regions). This is precisely the underlying reason behind the vulnerability of learning algorithms to blind-spot evasion.

Within this setting, evasion attacks can be conceived as a manipulation of the input samples $\vct x$ through a function $a(\vct x)$, which essentially introduces a deviation from the source distribution $p(\vct x, y)$. 
By denoting with $E_{a}(f) = \mathbb{E}_{(\vct x, y) \sim p } \ell( y, f(a(\vct x)))$ the error of the classifier $f$ on the manipulated samples, with $f^\prime$ the optimal (Bayesian) classifier on such manipulated data, we can compute the increase in the classification error of $f^\star$ on the manipulated data as the following:
\begin{equation}
\label{eq:error}
E_a(f^\star)-E(f^\star) = \underbrace{E_a(f^\prime)-E(f^\star)}_{\rm feature \; vulnerability} + \underbrace{E_a(f^\star) - E_a(f^\prime)}_{\rm learning \; vulnerability} \, .
\end{equation}
The first term is the increase in Bayesian error before and after the attack (which characterizes the vulnerability of the feature representation), while the second represents the classification error reducible by retraining on the attack samples (\ie, the vulnerability of the learning algorithm).

Under this interpretation, we can introduce a metric to assess the feature vulnerability quantitatively. 
To this end, we consider the so-called Bhattacharyya Coefficient (${\rm BC}$):
\begin{equation}
{\rm BC} = \int_{\vct x \in \set X} \sqrt{p_b(\vct x) p_m( \vct x)} d \vct x \in \{0,1\} \, .
\end{equation}
This coefficient essentially evaluates the overlapping between the distributions of benign $p_b$ and manipulated attack $p_m$ samples over the whole feature space $\set X$.
If the two distributions are the same, ${\rm BC}=1$, while if they are perfectly separated, ${\rm BC}=0$.
The convenient aspect of this metric is that it has a closed form for several known distributions; \eg, in the case of multivariate Gaussian distributions, it is given as ${\rm BC}= \exp (-D_B)$, where
\begin{equation}
\label{eq:db}
D_B = \frac{1}{8} (\vct \mu_b - \vct \mu_m)^\T \mat \Sigma^{-1} (\vct \mu_b - \vct \mu_m) 
+  \frac{1}{2} \log{\frac{{\rm det} \mat \Sigma }{\sqrt{ {\rm det} \mat \Sigma_b {\rm det} \mat \Sigma_m}}} \, ,
\end{equation}
and where $\Sigma = 0.5 (\Sigma_b + \Sigma_m)$, while $\mu_b$, $\mu_m$, $\Sigma_b$ and $\Sigma_m$ are the means and covariance matrices of benign and attack data, respectively.
To assess feature vulnerability, we use this expression for ${\rm BC}$, and exploit the well-known result that the Bayesian error is upper bounded by $\frac{1}{2} {\rm BC}$. One may indeed measure the difference between such value computed after and before the attack, which gives us an (approximate) indication of the increase in the Bayesian error induced by the attack, and thus, a quantitative measure of the feature vulnerability (\ie, of the first term in Eq.~\ref{eq:error}). Therefore, this coefficient can represent a useful measure that can be employed in practice to recognize the presence of feature vulnerabilities. We will use BC to discuss the efficacy of adversarial retraining in Section~\ref{sect:expeval:subsect:adv}.

\section{Experimental Evaluation}
\label{sect:expeval}
The experimental evaluation proposed in this paper is divided into two parts, which we describe in the following. 

	\myparagraph{Standard Accuracy Evaluation} In this evaluation, we tested the performances of FlashBuster against benign and malicious files in the wild by comparing them with the results attained by Hidost~\cite{srndic16-eurasip,hidost}. FlashBuster and Hidost were trained with a dataset of \emph{randomly chosen} malicious and benign SWF files, and they were tested against many previously unseen malicious and benign files. This experiment provided information on the \emph{general} performances attained by FlashBuster and Hidost regarding true and false positives. The goal of this evaluation was to ensure that the feature set we introduced obtained performances comparable to those attained by other publicly available state-of-the-art tools. Additionally, we tested the capability of FlashBuster to predict previously unseen attacks. To this end, we trained the system with data obtained before $2017$ and tested it against data obtained after the same year.  
	Finally, to ensure that FlashBuster could be reliably used also to detect new threats, we also performed an additional evaluation in which our system was tested against obfuscated and encrypted samples. 
		
	\myparagraph{Adversarial Evaluation} In this experiment (directly linked to the previous one), we evaluated the performances of FlashBuster against adversarial attacks performed according to a gradient descent strategy (see Section \ref{sect:evasion-attacks}). It is the first time that such evaluation has been carried out on Flash files and against Flash-based detection systems. Our goal was to understand the robustness of the features extracted by FlashBuster against adversarial modifications by employing possible defenses such as classifier retraining. Moreover, we provide a discussion about the efficacy of adversarial retraining as a possible defense to counteract adversarial attacks.  

In the following, we describe the dataset and the basic setup employed for all the experiments.

\subsection{Dataset}
\label{sect:expeval:subsect:dataset}

The dataset used for our experiments is composed of $5828$ files, $1593$ of which are malicious (an amount comparable to previous works \cite{overveldt12-raid,wressnegger16-dimva,srndic16-eurasip}) and $4235$ are benign (we chose this amount to work on balanced datasets~\cite{rossow12-sp}).
All malicious files, as well as part the benign ones, were retrieved from the VirusTotal service~\cite{virustotal}. Other benign files were retrieved from the DigitalCorpora repository~\cite{digitalcorpora}. Notably, for objective analysis, we only employed samples featuring code belonging to ASVM2 (as the previous version of the Virtual Machine employs different instructions and routines). Table~\ref{sect:expeval:subsect:dataset:tab:year} reports the distribution of the malicious samples by the date of the first submission to the service VirusTotal, together with the number of \emph{unique} CVEs that were correctly identified by the service\footnote{There may be further CVEs in the dataset that were not identified by VirusTotal. One CVE may apply to multiple files.}.

\begin{table}[!h]
	\begin{center}
		\caption{Distribution of malicious samples by the date of first submission to the VirusTotal service. The number of unique identified CVEs are reported for each year.}
		\label{sect:expeval:subsect:dataset:tab:year}
		\begin{tabular}
			{ |>{\bfseries}l  |c|c| }
			\hline
			\rowcolor[rgb]{.9,.9,1} \textbf{Year} &  \textbf{Num.Samples} & \textbf{Num. CVE} \\
			\hline
            $2018$ & $233$ & $3$ \\ \hline
            $2017$ & $149$ & $2$ \\ \hline
            $2016$ & $495$ & $2$ \\ \hline
            $2015$ & $453$ & $18$ \\ \hline
            $2014$ & $127$ & $9$ \\ \hline
            $2013$ & $45$ & $2$ \\ \hline
            $2012$ & $37$ & $4$ \\ \hline
            $2011$ & $18$ & $6$ \\ \hline
            $2010$ & $25$ & $2$ \\ \hline
            $2009$ & $9$ & $0$ \\ \hline
            $2009$ & $2$ & $0$ \\ \hline
			
		\end{tabular}
	\end{center}
\end{table}

The number of samples is well-balanced between $2015$ and $2018$, which were the years when Flash malware was employed the most by attackers. In particular, we report almost $30$ unique CVEs in $2015$ and $2016$. Overall, the employed dataset features a rather large diversity of vulnerabilities. This aspect is further confirmed by Table~\ref{sect:expeval:subsect:dataset:tab:families}, which shows the top-$10$ malware families contained in the dataset. We extracted the malware families by employing the popular tool \texttt{AvClass}~\cite{sebastian16-raid}, applied to the VirusTotal data. Note that most malware families are related to \emph{exploit-kits} (\eg, \texttt{neutrino}) that contain a specific Flash vulnerability.  This aspect shows that the analyzed samples belong to attacks that are consistently employed in the wild. Hence, we claim to have employed a dataset that is representative of the Flash-based malware ecosystem in the wild.  

\begin{table}[!h]
	\begin{center}
		\caption{Distribution of the top-$10$ malware families retrieved from the dataset of Flash-based malware.}
		\label{sect:expeval:subsect:dataset:tab:families}
		\begin{tabular}
			{ |>{\bfseries}l  |c| }
			\hline
			\rowcolor[rgb]{.9,.9,1} \textbf{Family} &  \textbf{Num. Samples} \\
			\hline
            \textbf{neutrino} & $114$ \\ \hline
            \textbf{axpergle} & $89$ \\ \hline
            \textbf{exkit} & $64$ \\ \hline
            \textbf{angler} & $45$ \\ \hline
            \textbf{swfexp} & $40$ \\ \hline
            \textbf{swif} & $38$ \\ \hline
            \textbf{lodabytor} & $35$ \\ \hline
            \textbf{swfdec} & $21$ \\ \hline
            \textbf{gwan} & $21$ \\ \hline
            \textbf{pubenush} & $15$ \\ \hline
			
		\end{tabular}
	\end{center}
\end{table}

\subsection{Basic Setup}
\label{sect:expeval:subsect:setup}

In this section, we describe the basic setup of the pre-processing, feature extractor, and classification modules for FlashBuster and Hidost. This setup is common to all the evaluations described in this section.

\subsubsection{Pre-Processing} 
\label{sect:expeval:subsect:setup:subsubsect:preproc}
Pre-processing was performed by FlashBuster and Hidost as follows: 

\begin{itemize}\item \textbf{FlashBuster.} As mentioned in Section~\ref{sect:architecture:subsect:parser}, the original \texttt{JPEXS} parser was modified to allow a faster analysis of multiple SWF files, as well as better integration with the other components of FlashBuster. All data related to tags and bytecodes were extracted and dumped into files, to allow for subsequent analyses by the other FlashBuster modules. The extraction time may vary from milliseconds to some minutes for very large files. 

\item \textbf{Hidost.} Hidost employed \texttt{SWFReTools}~\cite{swfretools}, a Java-based parser to analyze the structure of SWF files. After pre-processing, the analyzed files were added to a special cache to reduce the extraction times, which may vary from milliseconds to minutes.

\end{itemize}

\subsubsection{Feature Extraction.} 
\label{sect:expeval:subsect:setup:subsubsect:featextract}
We now provide some details about the feature extraction mechanisms employed by FlashBuster.
It is critical to point out that our experimental goal was \emph{not} developing the most accurate system possible, but performing an effective security evaluation under reasonable, practical constraints. 
In particular, there can be high differences in values among the employed features. Thus, attackers can create samples with abnormal values of certain features that would make the adversarial sample utterly different from the typical malware distribution (\eg, using specific functions thousands of times in the same method, while others are only used few times), thus achieving easy evasion. However, this evasion attempt is trivial, as anomalous values would make the sample easy to detect by using anomaly detectors.

To create a realistic situation for the attacker, we established an upper limit on each feature value in our dataset. For our experiments, we chose $10$ as a reasonable value that limits how often the same feature can be injected. We established this threshold empirically by performing statistics on how many features of the same type were typically contained in our dataset. Our findings showed that, while some samples may call the same API hundreds of times (especially in benign applications), the majority of the analyzed features were called (on average) less than $10$ times. Hence, the choice of this threshold represents the fact that it is unlikely that the majority of the features would appear more than $10$ times in a sample.
To confirm that limiting the feature values does not influence classification performances, we repeated our experiments with higher upper limits, without noticing significant differences in performances. 

\begin{figure*}[!htb]
    \centering
	\includegraphics[width=0.32\textwidth]{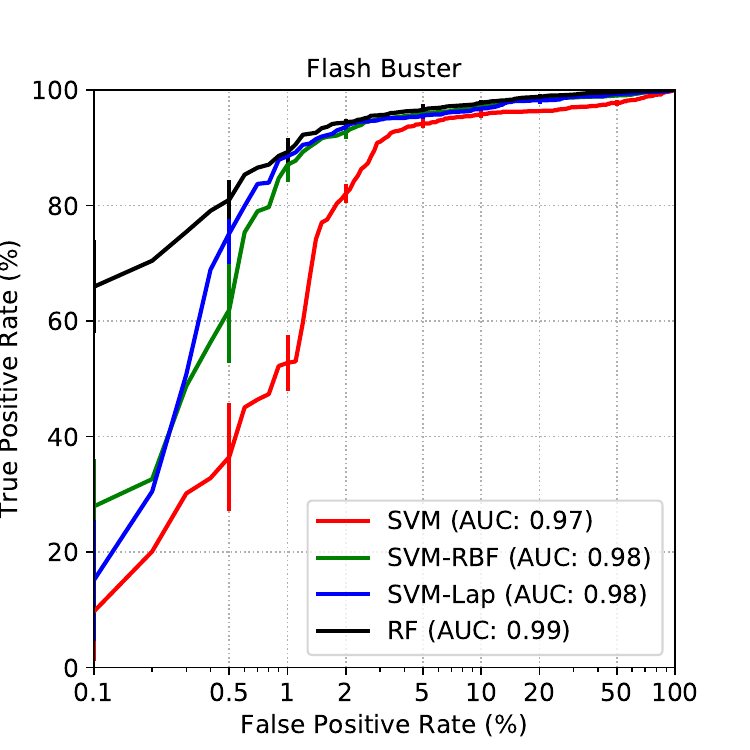}
	\includegraphics[width=0.32\textwidth]{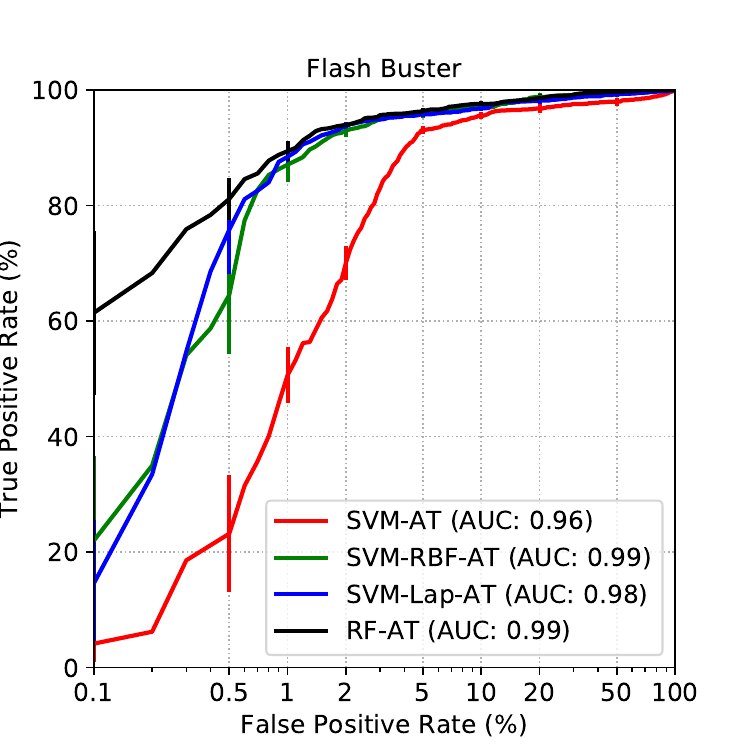}	
	\includegraphics[width=0.32\textwidth]{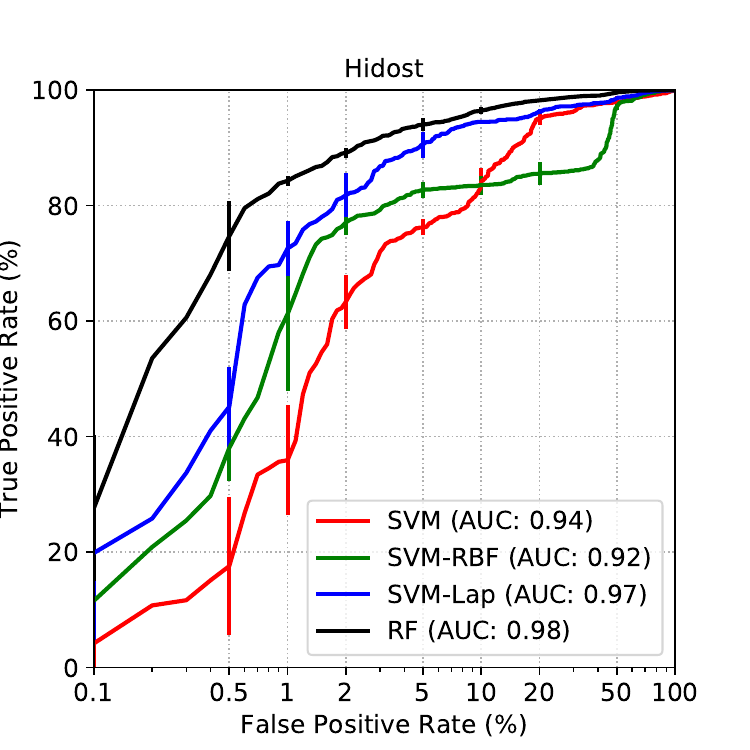}
	\caption{Average ROC curves and related AUCs obtained on $5$ train-test splits on FlashBuster (left - no adversarial retraining; center - with adversarial retraining) and Hidost (right).  
	\label{sect:expeval:subsect:std:fig:std_eval}
	}
\end{figure*}

\subsubsection{Classification and Training Procedures} 
\label{sect:expeval:subsect:setup:subsubsect:classification}
We used the popular machine-learning suite \texttt{scikit-learn}~\cite{scikit-learn}, which features the classifiers used in our evaluation,\footnote{Notably, Hidost was not released with a pre-trained classifier but only with the feature extraction module.} as well as \texttt{secml} to simulate evasion attacks against them~\cite{melis20-arxiv}.
All the performed evaluations (except the temporal evaluation and the one against obfuscated malware, which were carried out on an entirely different dataset) share the following elements:
 (a) The dataset was randomly split by considering 70\% of it as a training set and the remaining 30\% as a test set. The classifier hyperparameters were evaluated with a 5-fold cross-validation performed on the training set to optimize the true positive rate at 1\% false-positive rate. We repeated the whole procedure five times, in order to rule out possible biases related to specific train/test divisions. (b) We performed our tests on four classifiers: \emph{(i)} Random Forest (RF); \emph{(ii)} support vector machine with the linear kernel (SVM); \emph{(iii)} support vector machine with the Radial-Basis-Function kernel (SVM-RBF); \emph{(iv)} support vector machine with the Laplacian kernel (SVM-Lap). For support vector machines, we optimized the regularization hyperparameter $C$ and the kernel hyperparameter $\gamma$ (for kernel-based support vector machines) via cross-validation with $C, \gamma \in \{0.01, 0.5, 0.2, 0.1, 1, 2, 10, 20, 50, 100\}$. For random forests, we optimized the minimum number of samples required to split an internal node in $\{2, 5\}$, and either did not set any maximum depth or set it to 30 or 100. 
 
 \myparagraph{Adversarial Training (AT)} Under the same experimental setup, we retrained the same classifiers by adding the attacks generated against them to their training set. This procedure amounts to solving a minimax game known as \emph{adversarial training}~\cite{szegedy14-iclr,goodfellow15-iclr} in which the attacker maximizes the training loss of the classifier by manipulating the training points, while the classifier minimizes the same loss (computed on the manipulated samples) by adjusting its training parameters.
 We implemented an iterative version of this game in which, at each iteration, we first selected at random $500$ malware samples from the training set and optimized them to evade the target classifier. These samples were optimized by randomly-picking $\varepsilon \in \{1, 2, 5, 10, 20, 50, 100 \}$. Then, we added such adversarial samples to the training set and retrained the target classifier on the augmented training data. We terminated this procedure after $10$ iterations, or earlier if convergence to a stable solution was observed.
 We denote the corresponding robust classifiers with the suffix AT, \ie, SVM-AT, SVM-RBF-AT, SVM-Lap-AT, and RF-AT.
   
\subsection{Standard Accuracy Evaluation}
\label{sect:expeval:subsect:std}

The standard accuracy evaluation was performed by following the criteria described in Section~\ref{sect:expeval:subsect:setup:subsubsect:classification}. For both FlashBuster and Hidost, and for each of the four classifiers tested, we calculated the average (among its splits) Receiving Operating Characteristic (ROC), which reports the detection rate (\ie, the fraction of correctly-detected malware) at different false-positive rates (FPR, \ie, the fraction of misclassified benign samples).

Results are shown in Figure~\ref{sect:expeval:subsect:std:fig:std_eval}. The leftmost plot shows that FlashBuster detects more than $90\%$ malware samples at $1\%$ FPR, while at $5\%$ the detection rate is higher than $95\%$. Non-linear models such as Random Forests and SVMs with the Laplacian kernel perform better than their linear counterparts. The middle plot reports the results obtained by retraining the classifiers with adversarial attacks (described more in detail in Section~\ref{sect:expeval:subsect:adv}). The attained results show that, despite adding artificially-generated samples to the training set, the detection rate of the classifiers does not substantially change. We discuss these aspects more in detail in the next section.

The rightmost plot shows the results attained by Hidost. FlashBuster performs better than Hidost with every classifier, with a difference in detection rates between $10$\% and $20$\%, depending on the selected classifier. We interpret these results with the fact that Hidost employs only structural features without analyzing the bytecode embedded in Flash applications. Some SWF files may indeed employ very similar structures, but extremely different codes. Hence, such files could be difficult to detect with purely structural approaches. This aspect has also been encountered in PDF files, where purely structural-based methods may fail with malicious files with many employed objects~\cite{maiorca19-csur}. 

Overall, the attained curves show that FlashBuster can be effectively used as a detector of malicious files, outperforming competing state-of-the-art approaches.

\subsubsection{Temporal Evaluation}
\label{sect:expeval:subsect:std:subsubsect:temporal}
We evaluated the ability of FlashBuster to predict previously-unseen attacks. To this end, we trained the system by only using samples whose first submission date to the \texttt{VirusTotal} service was previous to $2017$ (a total of $1211$ malicious samples), plus all the benign files in the dataset. The test set was therefore made of those malicious samples released in $2017$ and $2018$ ($383$ samples). We used all the classifiers of the previous experiments (along with retrained classifiers), and the parameters of the classifiers were evaluated with a $5$-fold cross-validation performed on the training set. 
 
Table~\ref{sect:expeval:subsect:temp:tab:results} shows the results for this evaluation. It is possible to see that RF and SVM-Lap report very high accuracy values (almost $90$\%) on the test set. This result is also in-line to what showed in Figure~\ref{sect:expeval:subsect:std:fig:std_eval}. It is interesting to point out that linear classifiers, which are the ones to perform worst against malware in the wild, provide good accuracy ($85$\%) on unseen attacks. SVM-RBF classifiers are the ones to perform worst with a $75$\% of accuracy. Notably, the linear SVM significantly improves its performances after having been retrained with adversarial samples. We speculate that this behavior is due to similarities between the adversarial generated samples and the novel samples belonging to the test set. Overall, all models prove to be reliable at detecting previously-unseen malware, showing that FlashBuster can be effectively used for this task.

\begin{table}[!h]
	\begin{center}
		\caption{Accuracy performances on five classifiers on a test set composed of data released from $2017$. Each classifier has been trained with data released before $2017$. We also report the performances of retrained classifiers.}
		\label{sect:expeval:subsect:temp:tab:results}
		\begin{tabular}
			{ |>{\bfseries}l  |c|c| }
			\hline
			\rowcolor[rgb]{.9,.9,1} \textbf{Classifier} &  \textbf{Acc.} & \textbf{Acc. (Retrain)} \\
			\hline
			RF & $88\%$ & $89\%$ \\ \hline
			SVM Laplacian & $88\%$ & $89\%$ \\ \hline
			SVM Linear &  $85\%$ & $89\%$\\ \hline
			SVM RBF &  $75\%$ & $87\%$\\ \hline
		\end{tabular}
	\end{center}
\end{table}

\subsubsection{Evaluation against Obfuscated Malware}
\label{sect:expeval:subsect:std:subsubsect:obfuscation}

To test the detection capabilities of FlashBuster against obfuscated samples, we generated two additional datasets of  malicious samples by using the popular obfuscation tool \texttt{DoSWF}~\cite{doSWF}, starting from the original malicious dataset. Notably, as \texttt{DoSWF} could not obfuscate all the samples (due to technical limitations of the tool), the two generated datasets featured fewer samples than the starting one. The datasets were organized as follows:

\begin{itemize}
    \item \textbf{Obfuscated.} This dataset was generated by replacing variable names and by performing changes to the control flow graph without modifying the file semantics. After obfuscation, we obtained $1278$ obfuscated samples.
    \item \textbf{Encrypted.} This dataset was generated by making each file be dynamically generated at runtime. Typically, encryption is employed to dynamically load files in memory through customized routines introduced by the obfuscator. After encryption, we obtained $1466$ samples.  
\end{itemize}

We used a training set composed of all malicious and benign files employed during the experiments reported in Section~\ref{sect:expeval:subsect:std}. We then tested the various classifiers of FlashBuster (including the ones featuring retrained data) on the two datasets. Results are reported in Tables~\ref{sect:expeval:subsect:obfusc:tab:obfuscated} and~\ref{sect:expeval:subsect:obfusc:tab:encrypted}, and show that FlashBuster is able to detect almost all obfuscated and encrypted samples in the wild. 

\begin{table}[!h]
	\begin{center}
		\caption{Accuracy performances (in percentage) on five classifiers on a test set composed of obfuscated data. We also report the performances of retrained classifiers.}
		\label{sect:expeval:subsect:obfusc:tab:obfuscated}
		\begin{tabular}
			{ |>{\bfseries}l  |c|c| }
			\hline
			\rowcolor[rgb]{.9,.9,1} \textbf{Classifier} &  \textbf{Acc.} & \textbf{Acc. (Retrain)} \\
			\hline
			RF & $97\%$ & $98\%$ \\ \hline
			SVM RBF &  $95\%$ & $96\%$\\ \hline
			SVM Laplacian & $94\%$ & $94\%$ \\ \hline
			SVM Linear &  $93\%$ & $92\%$\\ \hline

		\end{tabular}
	\end{center}
\end{table}

This result may seem rather surprising because obfuscated malware should be significantly different from the original one. However, there are two explanations to this effect: \emph{(i)} The features employed by FlashBuster are related to system-API calls, and the impact of obfuscation on these features is rather limited. In some cases, encryption may add other system API-based routines, which may increase the maliciousness of the file for the classifier; \emph{(ii)} \texttt{DoSWF} has been often used as a way to conceal malware in the wild. Hence, we speculate that obfuscated and encrypted malware was already in the employed training set. As a consequence, the classifier learned the characteristics of obfuscated malware and recognized obfuscated attacks due to their characteristics (an aspect also observed in a very recent work~\cite{mantovani-ndss20}). 

\begin{table}[!h]
	\begin{center}
		\caption{Accuracy performances (in percentage) on five classifiers on a test set composed of encrypted data. We also report the performances of retrained classifiers.}
		\label{sect:expeval:subsect:obfusc:tab:encrypted}
		\begin{tabular}
			{ |>{\bfseries}l  |c|c| }
			\hline
			\rowcolor[rgb]{.9,.9,1} \textbf{Classifier} &  \textbf{Acc.} & \textbf{Acc. (Retrain)} \\
			\hline
			RF & $100\%$ & $100\%$ \\ \hline
			SVM RBF &  $100\%$ & $100\%$\\ \hline
			SVM Laplacian & $94\%$ & $94\%$ \\ \hline
			SVM Linear &  $100\%$ & $100\%$\\ \hline

		\end{tabular}
	\end{center}
\end{table}

\subsection{Adversarial Evaluation}
\label{sect:expeval:subsect:adv}

The adversarial evaluation aimed to assess the performance of the classifiers employed in the previous experiment after the gradient descent attacks described in Section~\ref{sect:evasion-attacks}. In this case, we evaluated how the detection rate (at 1\% FPR) of the classifiers decreases against an increasing number of injected features $\varepsilon$. According to this security evaluation procedure, a classifier is said to be more robust if its detection rate decreases more gracefully~\cite{biggio18}.

\begin{figure*}[!htb]
    \centering
	\includegraphics[width=0.32\textwidth]{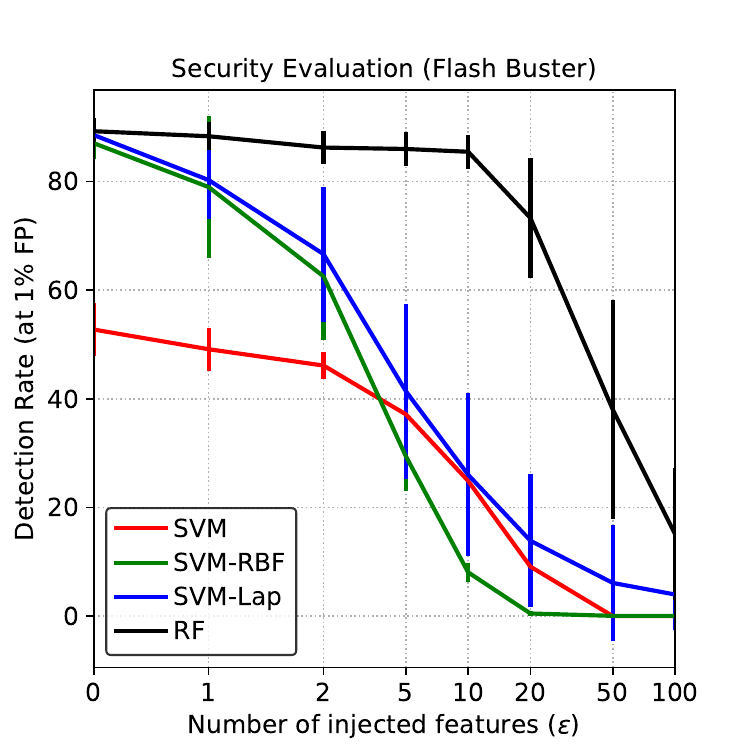}
	\includegraphics[width=0.32\textwidth]{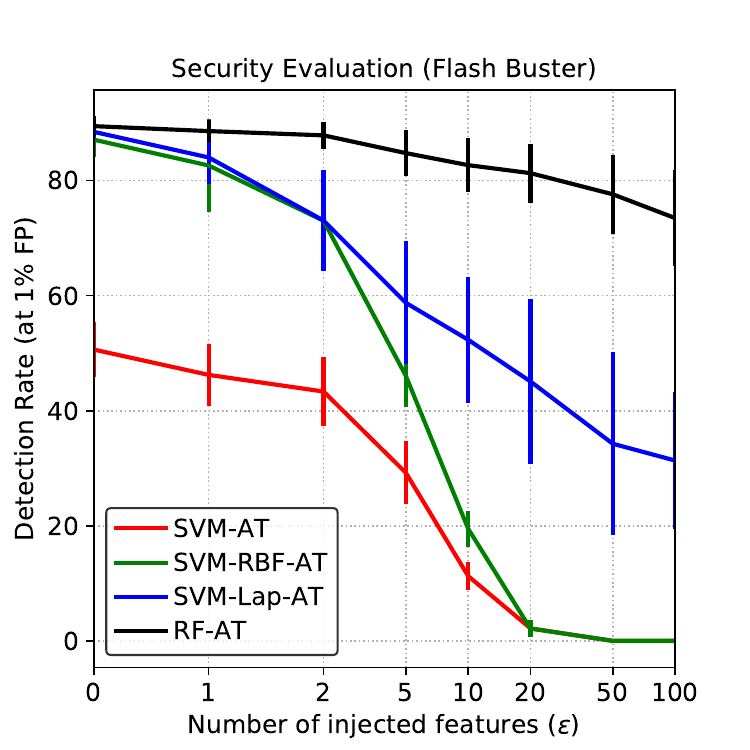}	
	\caption{Security evaluation curves on FlashBuster. On the x-axis, we report the number of changes to the feature values is reported. On the y-axis, we report the decrease of the detection rate as more features are modified. The left-side curves are related to the classifiers that were not retrained with adversarial attacks, while the right-side ones concern the classifiers under adversarial retraining.} 
	\label{sect:expeval:subsect:adv:fig:sec-eval-flash-buster}
\end{figure*}

It is important to observe that RF classifiers are not differentiable, so we can not use our gradient-based attack directly to evaluate their security. We thus attacked our RF classifiers by first optimizing the attacks against all the SVM-based classifiers (used as surrogate models), and then evaluating whether such attacks \emph{transfer} correctly to the RF classifiers (Limited Knowledge - surrogate models). In practice, we found that the attacks exhibiting the highest success against the RF classifiers were those optimized against the SVM-Lap models.

The leftmost plot in Figure~\ref{sect:expeval:subsect:adv:fig:sec-eval-flash-buster} provides the results of the evaluation when the classifiers are not retrained with adversarial attacks. While all SVM-based classifiers are completely evaded after $\varepsilon = 100$ changes, Random Forests can still detect $20\%$ of the attacks. This effect may be present because we are using a surrogate model to attack the RF classifiers rather than directly optimizing the attack against them.
For this reason, we can not state with certainty that RFs are \emph{generally} more secure; indeed, a more powerful white-box attack as that in~\cite{kantchelian16-icml} may enable evading them with higher probability. We thus leave a more detailed investigation of this aspect to future work.

The rightmost plot in Figure~\ref{sect:expeval:subsect:adv:fig:sec-eval-flash-buster} shows the security evaluation curves obtained by retraining the classifiers with samples generated through adversarial attacks. While the retraining strategy essentially brings no additional robustness for SVM and SVM-RBF classifiers, we point out a significant increment of robustness in SVM-Lap and RF classifiers. 
This effect may be due to the decision function learned by these models, which may be better shaped to counter the presence of $\ell_1$-norm adversarial attacks (as those simulated in this work, where the attack algorithm manipulates only a few relevant features due to the presence of the sparse constraint $\| \vct x^\prime - \vct x\|_1 \leq \varepsilon$)~\cite{russu16-aisec}.

\begin{figure*}[!htb]
    \centering
	\includegraphics[width=0.24\textwidth]{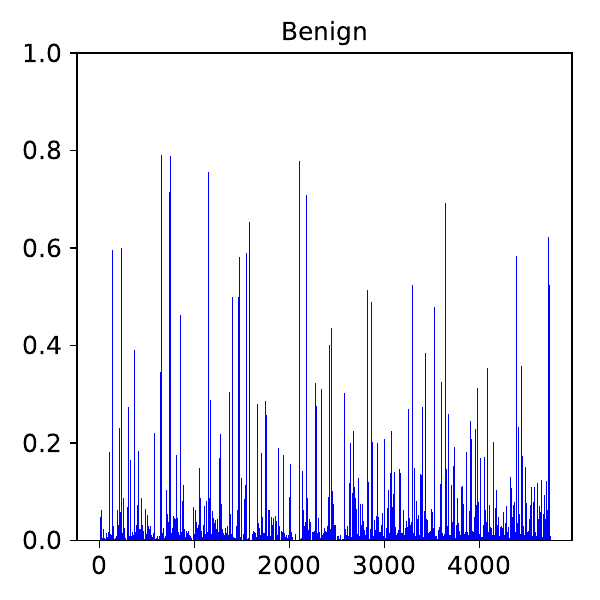}
	\includegraphics[width=0.24\textwidth]{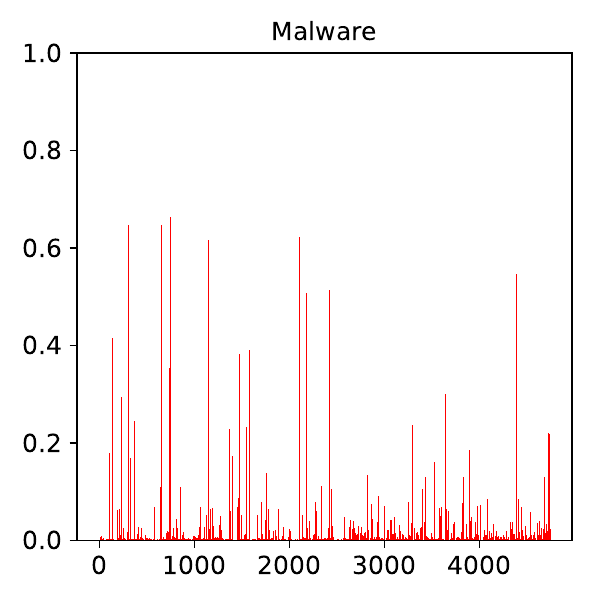}
	\includegraphics[width=0.24\textwidth]{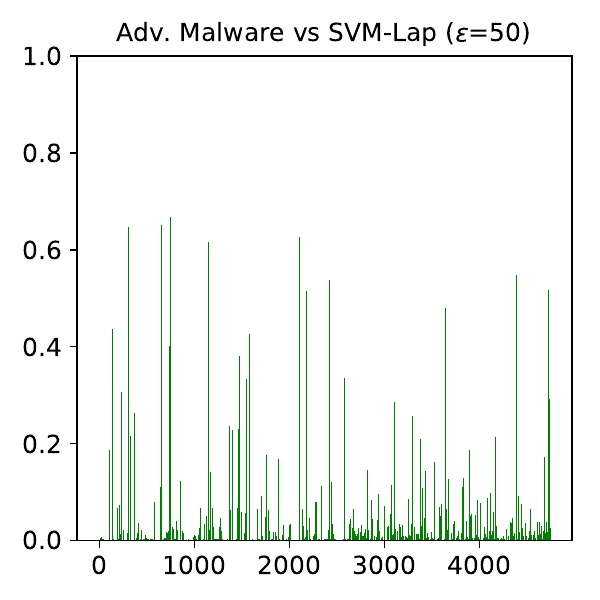}
	\includegraphics[width=0.24\textwidth]{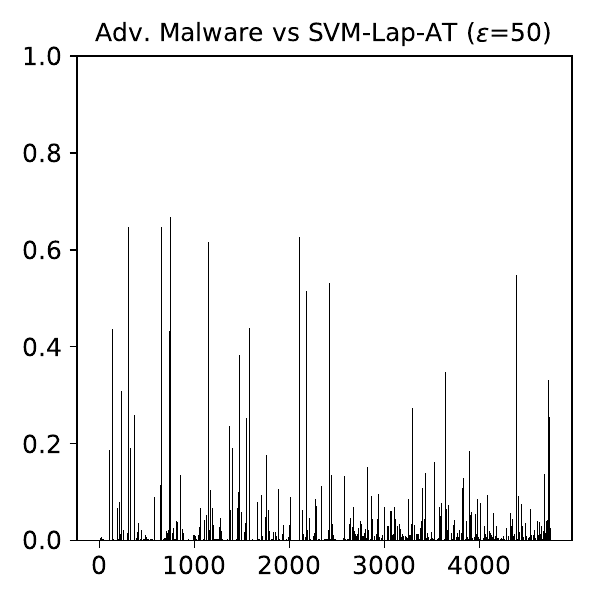}
	\caption{Average feature values for one test set in which malicious samples (red) were modified with our gradient-based attack ($\varepsilon$ is the number of changes) against SVM-Lap (green) and SVM-Lap-AT (black). The Bhattacharyya coefficient $BC$ values between these malicious (red, green and black) and benign (blue) samples are reported in Figure~\ref{sect:expeval:subsect:adv_eval:fig:feat_distr2}.}
	\label{sect:expeval:subsect:adv_eval:fig:feat_distr1}
\end{figure*}

\begin{figure*}[!htb]
    \centering
	\includegraphics[width=0.8\textwidth]{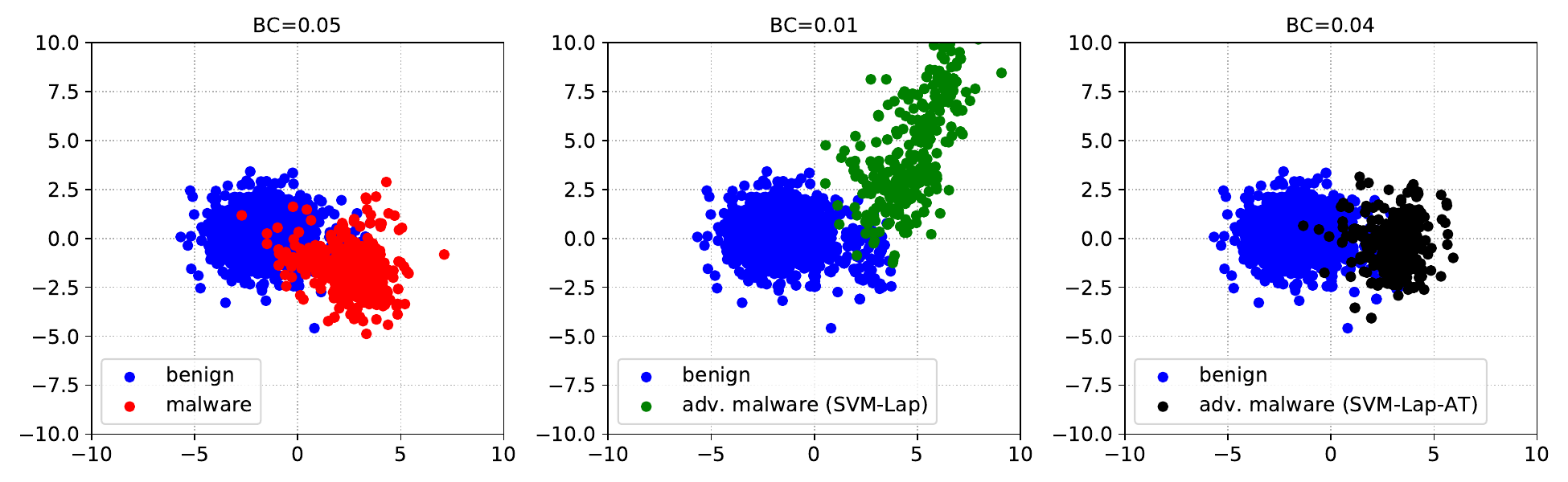}
	\caption{Two-dimensional LDA projections of benign samples against malware (leftmost plot), adversarial malware targeting SVM-Lap (middle plot) and adversarial malware targeting SVM-Lap-AT (rightmost plot), along with the corresponding BC values.}
	\label{sect:expeval:subsect:adv_eval:fig:feat_distr2}
\end{figure*}

\myparagraph{Discussion on Adversarial Training}
We investigate here why adversarial training is only slightly improving robustness to adversarial attacks in this case. 
To this end, in Figure~\ref{sect:expeval:subsect:adv_eval:fig:feat_distr1} we report the average feature values (\ie, the centroid) for benign and malware data (normalized in $[0,1]$ after division by 10), extrapolated from one test set, and the average feature values for adversarial malware samples optimized against SVM-Lap and SVM-Lap-AT.
For the same data, we also report in Figure~\ref{sect:expeval:subsect:adv_eval:fig:feat_distr2} the variation of the related Bhattacharyya coefficient $BC$ (described in Section~\ref{sect:vuln}) computed between the malicious and benign distributions after projection on a two-dimensional space via Linear Discriminant Analysis (LDA). 

From Figure~\ref{sect:expeval:subsect:adv_eval:fig:feat_distr1}, it may appear that the centroids of malware and adversarial malware are not that different. This phenomenon is correct since we are only considering the injection of $\varepsilon=50$ features in each sample. However, the two-dimensional projection elaborated in Figure~\ref{sect:expeval:subsect:adv_eval:fig:feat_distr2} unveils an interesting phenomenon. First of all, attacking SVM-Lap and SVM-Lap-AT decreased the initial BC value. This effect is surprising since it means that the attacks are effective even though the Bayesian error (distribution overlap) might decrease. In other words, both attacks evade detection by placing the adversarial malware samples in blind spots, without mimicking the characteristics of the benign distribution. 
Nevertheless, adversarial malware optimized against the non-robust SVM-Lap classifier is much more separable than the adversarial malware optimized against the robust SVM-Lap-AT classifier from the benign data (cf. the BC values in the middle and rightmost plots).
This means that adversarial training, even after several rounds of retraining and addition of up to $5,000$ adversarial malware samples in the training set, is capable of enforcing the attack samples to better reproduce the behavior of legitimate samples to be successful.
However, even though the classifier has eventually learned a more robust decision surface in our case, this is not sufficient to stop practical attacks in this specific domain. In fact, by injecting more features, the attacker may be able to even better mimic the benign distribution and evade detection with higher probability. 
In addition, adversarial training is a very costly procedure in terms of computational complexity, both in space and time, especially in high-dimensional spaces like in our case (FlashBuster generates more than $4,000$ distinct features), as it requires generating a huge number of samples (potentially exponential in the number of features) before being slightly effective.
Accordingly, robust methods based on regularization rather than data augmentation may be better suited to our case~\cite{demontis17-tdsc,lyu15-icdm,simon18,ross18}.

To summarize, the attained results show that, although the function was retrained against the attacks to reduce the \emph{learning} vulnerability, the \emph{feature} vulnerability related to the employed static features could not be reduced by merely retraining the classifier. In more detail, the problem here is that \emph{none of the employed static features is likely to appear in malware much more frequently than in benign data}, \ie, there is no \emph{invariant} feature that characterizes malware uniquely from benign data (and that can not be removed)~\cite{tong17-arxiv}.
Such lack of invariance means that, in principle, it is possible to create a malicious sample that is \emph{indistinguishable} from benign ones (even by only injecting content to the input SWF file) and, thus, additional features are required to detect adversarial SWF malware examples correctly.

\section{Summary of Results and Discussion}
\label{sect:overall}
In this Section, we provide a summary of the results attained during the experimental Section, along with a brief discussion on each point.

\begin{itemize}
    \item \textbf{Detection of malware in the wild}. FlashBuster can effectively detect malware in the wild. In particular, it features significantly higher performances than Hidost, a valid state-of-the-art tool for SWF malware detection. Using structural- and content-based features provides a more reliable detection against malware with structures similar to benign files. Overall, we demonstrated that our system could be validly used as a reference for further experiments related to robustness.
    \item \textbf{Detection of previously unseen and encrypted malware}. FlashBuster provides very good accuracy at detecting samples released after training data. Moreover, FlashBuster can detect obfuscated and encrypted (by a popular off-the-shelf tool) malware. In this last case, the distribution of training data is critical to ensure that the malware could be effectively detected. In particular, if the distribution of encrypted malware significantly differs from the one of normal malicious or benign files, it is likely that the system would not be able to detect it. Conversely, if samples obfuscated with similar techniques are present in the training set, the system's performances can be extremely high.
    \item \textbf{Evasion attacks against classifiers}. When no retraining is employed, gradient-based attacks can completely evade the detection of all classifiers after a reasonable number of changes. 
    \item \textbf{The role of adversarial training}. Adversarial training can slightly improve robustness in some cases, but not to a satisfying extent for this application. The main issue here is that defending the classifier remains useless if the adopted feature representation is vulnerable (\ie, the attacks are able to mimic the benign feature values). Moreover, adversarial training remains very computationally demanding, hindering both its effectiveness and scalability in high-dimensional domains such as that considered in this work.
    
\end{itemize}

\section{Conclusions and Future Work}
\label{sect:conclusions}

In this paper, we proposed a security evaluation of static malicious SWF file detectors by introducing FlashBuster, a system that combines structural and content-based information to perform an accurate analysis of such attacks. In particular, we demonstrated that FlashBuster could attain improved performances (in terms of detection rate) compared to other state-of-the-art tools. Moreover, we showed that it could be effectively employed to detect previously unseen, obfuscated, and encrypted attacks.  
The proposed security evaluation showed an intrinsic vulnerability of the static features used by SWF detectors. In particular, by using gradient descent attacks, we demonstrated how even retraining strategies were not always effective at ensuring robustness. More specifically, we measured and showed how gradient descent attacks made samples more similar to their benign counterparts, thus making adversarial retraining inefficient in some cases.
We plan to improve and solve some of the system limitations in future work: for example, reducing its dependence on \texttt{JPEXS}, whose possible failures could compromise the whole file analysis. We also plan to perform more experiments on SWF files that are obfuscated with off-the-shelf tools in order to evaluate the resilience of FlashBuster against them.

In general, our claim for future research is that focusing on improving the classifier decision function can be effective only if the employed features are intrinsically robust, \ie, there should be specific features that are \emph{truly characteristic} of malicious behavior and that cannot be mimicked in benign files (or whose mimicking would require a significant effort from the attacker's side). For example, the maliciousness of a feature could be better pointed out by considering the context in which it was found. There may be API calls that could be solely used for malicious purposes if they operate after (or before) other API calls, with specific parameters, or when the program is in a determined state of execution. We believe that this research direction will be useful not only for Flash malware detection but also for the other malware detection systems.

\section*{Acknowledgments}
This work was partly supported by the project PON AIM Research and Innovation 2014-2020 - Attraction and International Mobility, funded by the Italian Ministry of Education, University and Research; by the PRIN 2017 project RexLearn, funded by the Italian Ministry of Education, University and Research (grant no. 2017TWNMH2); and by the EU H2020 project ALOHA, under the European Union’s Horizon 2020 research and innovation programme (grant no. 780788). The authors would like to thank Maria Elena Chiappe, Denis Ugarte and Michele Scalas for their contributions to FlashBuster. 

\bibliographystyle{model1-num-names}
\balance
\bibliography{biblio,malwDB}

\end{document}